\documentclass{aa}
\usepackage{epsfig}
\usepackage{graphicx}
\usepackage{natbib}
\usepackage{wasysym}
\usepackage{txfonts}

\bibpunct{(}{)}{;}{a}{}{,}

    \def\void#1{{}}

    \def\etal{{\it et al.\ }}
    \def\be{\begin{enumerate}}
    \def\ee{\end{enumerate}}
    \def\bi{\begin{itemize}}
    \def\ei{\end{itemize}}
    
    \def\dg{$^{\circ}$}

    \def\~{$\sim$}
    \def\={$\simeq$}
    
    \def\H0{H_{0} \/$}
    \def\h0{H_{0} \/$}
    \def\h-1{$h^{-1} \/$}

\newcommand{\mincir}{\raise -2.truept\hbox{\rlap{\hbox{$\sim$}}\raise5.truept
\hbox{$<$}\ }}
\newcommand{\magcir}{\raise -2.truept\hbox{\rlap{\hbox{$\sim$}}\raise5.truept
\hbox{$>$}\ }}
\newcommand{\minmag}{\raise-2.truept\hbox{\rlap{\hbox{$<$}}\raise
6.truept\hbox
{$>$}\ }}
\def\lsimeq{{_<\atop^{\sim}}}
\def\gsimeq{{_>\atop^{\sim}}}

\begin{document}
\title{The ATESP 5 GHz radio survey.\\ 
III. 4.8, 8.6 and 19 GHz follow-up observations of radio galaxies}

\author{I. Prandoni\inst{1} \and H.R. de~Ruiter\inst{1} \and 
R. Ricci\inst{1,2} \and P. Parma\inst{1} \and L. Gregorini\inst {3,1}
\and R.D. Ekers\inst{4}}

\offprints{I. Prandoni, \email{i.prandoni@ira.inaf.it}}
\institute{ INAF - Istituto di Radioastronomia, Via Gobetti~101, I-40129 
Bologna, Italy
\and Department of Physics and Astronomy, University of Calgary, 2500 
University Drive NW, Calgary, AB T2N 1N4, Canada  
\and Dipartimento di Astronomia, Universit\`{a} di Bologna, via 
Ranzani~1, I-40127 Bologna, Italy 
\and Australia Telescope National Facility, CSIRO,
PO Box 76, Epping NSW 1710, Australia
}

\date{Received -; Accepted -}
\titlerunning{The ATESP 5~GHz radio survey. III}
\authorrunning{I. Prandoni \etal}

\abstract
{}
{Physical and evolutionary properties of the sub-mJy radio population 
are not entirely known. 
The radio/optical analysis of the ATESP 5~GHz sample 
has revealed a significant class of compact flat/inverted radio-spectrum 
sources associated to early-type galaxies up to redshift $2$. 
Such sources are most plausibly triggered by an AGN, but 
their observational properties are not entirely consistent with those of 
standard radio galaxy populations. 
In the present work we aim at a better understanding of the radio 
spectra of such sources and ultimately of the nature of AGNs at sub-mJy
flux levels. In particular we are interested in assessing whether the AGN
component of the sub--mJy population is more related to
efficiently accreting systems - like radio-intermediate/quiet quasars - or to
systems with low accretion rates - like e.g. FRI radio galaxies - or to
low radiative efficiency accretion flows - like e.g. 
ADAF. 
}  
{We used the ATCA to get multi-frequency
(4.8, 8.6 and 19 GHz) quasi-simultaneous observations for a representative 
sub-sample of ATESP radio sources associated with early-type galaxies 
(26 objects with $S>0.6$ mJy). This can 
give us insight into the accretion/radiative mechanism that is at work, 
since different regimes display different spectral signatures in 
the radio domain.
} 
{From the analysis of the radio spectra, we find that our sources are 
most probably jet-dominated systems. ADAF models are ruled out by the high 
frequency data, while ADAF+jet scenarios are still consistent with 
flat/moderately
inverted-spectrum sources, but are not required to explain the data. 
We compared our sample with high ($\gsimeq 20$ GHz) frequency 
selected surveys, finding spectral properties very similar to the ones of
much brighter ($S>500$ mJy) radio galaxies extracted from the 
Massardi \etal (2008) sample. 
Linear sizes of ATESP 5~GHz sources associated with early type galaxies 
are also often consistent with the ones of brighter B2 and 3C radio galaxies, 
with possibly a very compact component that could be associated at least in 
part to (obscured) radio-quiet quasar-like objects and/or 
low power BL Lacs.}
{}


\maketitle

\section{Introduction}
\label{sec:introduction}
After many years of studies, AGNs are now recognised to contribute 
significantly at radio fluxes below 1 milliJy (mJy).
Multi-wavelength studies of deep radio fields show that star-forming
galaxies dominate at microJy ($\mu$Jy) levels 
\citep[Richards~\etal][]{Ric99}, 
while radio sources associated with
early--type galaxies and plausibly triggered by AGNs, are the most
significant source component ($\sim 60-70\%$ of the total) 
at higher flux densities ($>100-200$ $\mu$Jy)
with a further 10\% contribution from
broad-/narrow-line AGNs (see e.g. Gruppioni~\etal~\citealp{Gru99}; 
Georgakakis~\etal~\citealp{Geo99}; Magliocchetti~\etal~\citealp{Mag00};
Prandoni~\etal~\citealp{Pra01b}; Muxlow~\etal~\citealp{Mux05}; 
Afonso~\etal~\citealp{Afo06}; Mignano ~\etal~\citealp{Mig08}; 
Smolcic~\etal~\citealp{Smo08}).
This somehow unexpected presence of large numbers of AGN--type sources at the
sub-mJy level has given a new and interesting scientific perspective to the
study of deep radio fields, since a better 
understanding of the physical and evolutionary properties of 
such low/intermediate power AGNs may have important implications for the 
determination of the black-hole-accretion history
of the Universe as derived from radio-selected samples. 
Of particular interest is the possibility of assessing whether the AGN
component of the sub--mJy population is more related to
efficiently accreting systems - like radio-intermediate/quiet quasars - or to
systems with very low accretion rates - like e.g. FRI radio galaxies 
\cite[Fanaroff \& Riley][]{Fan74}. The
latter scenario (radio mode) is supported by the presence of many optically
inactive early type galaxies among the sub--mJy radio sources. The
quasar mode scenario, on the other hand, may be supported by the large 
number of so-called
radio-intermediate quasars observed at mJy levels 
\cite[see e.g. Lacy \etal][]{Lac01} and by the modelling work of Jarvis \& 
Rawlings \cite{Jar04}, 
who predict a significant contribution of radio-quiet quasars at 
sub-mJy levels.
Another issue is represented by the possible role played at low radio fluxes 
by low 
radiation efficiency accretion mechanisms, associated to optically thin discs,
such as the so-called {\it advection dominated accretion flows} (ADAF) and 
modifications (ADIOS, CDAF, etc; see Narayan \& Yi~\citealp{Nar94}; Quataert \&
Narayan~\citealp{Qua99}; Abramowitcz~\etal~\citealp{Abr02})

One especially suited radio sample to study the phenomenon of low-luminosity 
nuclear activity, possibly related to low radiation/accretion processes and/or 
radio-intermediate/quiet QSOs, is the so-called ATESP 5~GHz survey, 
carried out with the Australia Telescope Compact Array (ATCA) by
Prandoni~\etal~\cite{Pra06}, which covers 
a $2\times 0.5$ sq.~degr. region of the wider original 1.4 GHz ATESP 
survey \citep[Prandoni~\etal][]{Pra00a,Pra00b,Pra01a}. The ATESP 5~GHz survey 
consists of 131 radio sources with flux densities larger than 
$S\sim 0.4$ mJy detected either at 1.4~GHz or 5~GHz or both. 
Interestingly, the analysis of the 1.4 and 5~GHz ATESP surveys has revealed 
a significant flattening of the source radio spectra going from mJy to 
sub-mJy flux levels (Prandoni~\etal\citealp{Pra06}). Such flattening is mostly
associated to early-type galaxies as demonstrated by deep ($R<25$) 
UBVRIJK multi-colour imaging. At the flux limit of the sample star-forming 
galaxies are starting to appear, but are not yet the dominant population: 
$\sim 14\%$ of the sources are 
identified with broad/narrow-line AGNs and another $\sim 64\%$ is 
identified with 
early-type galaxies, most probably triggered by an AGN 
\cite[see Mignano~\etal][]{Mig08}.
The absence of emission lines in the latter together with low 
radio luminosities (typically $10^{22-25}$ W/Hz) 
suggests that these are FRI radio galaxies. However, $>60\% $ of them 
show flat ($\alpha >-0.5$, where $S\sim \nu^{\alpha}$) and/or inverted 
($\alpha >0$) spectra and rather compact linear sizes ($d<10-30$ kpc, 
\citealp[see Mignano~\etal][]{Mig08}). Both compactness and spectral shape 
suggest a core emission with strong synchrotron or free-free self-absorption. 
Such sources are known to exist among FRI radio galaxies, but 
they are relatively rare. It is therefore important to confirm the existence
of such a large class of flat/inverted spectrum objects through 
simultaneous radio observations at different radio wavelengths. 

If real, these sources may represent a composite class of objects very 
similar to the so-called low power ($P_{408 MHz}<10^{25.5}$ W/Hz) 
compact ($<10$ kpc) - LPC - radio sources studied by 
Giroletti~\etal~\citep{Gir05}. LPC host galaxies do not show signatures of
strong nuclear activity in the optical (and X-ray) bands, and preliminary 
results
indicate that multiple causes can produce LPC sources: geometrical-relativistic
effects (low power BL-Lacertae objects), youth (GPS-like sources), 
instabilities in the jets, frustration by a denser than average ISM and a 
premature end of nuclear 
activity (sources characterised by low accretion/radiative efficiency, i.e. 
ADAF/ADIOS systems). Some difficulties remain however. First, if the 
objects are young GPS sources, much higher ($>10^{25}$ W/Hz)
radio luminosities are expected;
second, if they are old ADAF/ADIOS sources, the expected luminosities are
much lower ($<10 ^{21}$ W/Hz, Doi \etal \citealp{Doi05}). 
In the latter case there may yet be different solutions, as for example if 
the ADAF source 
coexists with a radio jet, as was suggested by Doi \etal \citep{Doi05} in the 
case of a few low luminosity AGNs 
\cite[see also the ADAF-jet model of Falcke \& Biermann][]{Fal99}.

Another possibility is that we are dealing with 
radio-quiet/intermediate QSOs, where the activity in the optical band is 
obscured by dust in the galaxy. This scenario may be supported by 
the results reported by Kl\"ockner~\etal~\citep{Klo09}, who have recently 
observed with the EVN 11 $z>2$ radio-intermediate obscured quasars 
(several with flat-spectrum), detecting seven of them. The detected radio emission 
accounts for 30-100\% of the entire source flux
density, and the physical extent of this emission is $\lsimeq 150$ pc. 
The missing flux implies the likely existence of radio jets of physical sizes 
between $\gsimeq 150$ pc and $\lsimeq 40$ kpc. 

The aim of the present paper is a better understanding of the 
physical properties of the AGNs associated to early-type 
galaxies in the ATESP 5~GHz sample. The study of the radio spectra, based 
on multi-wavelength (simultaneous) observations, is essential in this respect, 
since different source types ({\it e.g.} GPS vs BL Lac) and/or accretion 
scenarios ({\it e.g.} jet-dominated vs ADAF/ADIOS-dominated) lead to different 
shapes of the radio spectrum. Particularly useful is the availability of 
high radio frequency information, since, for example the location of the 
frequency at which the radio spectrum peaks may be an indication of the 
presence or absence of outflows. A pure ADAF model without outflow is 
expected to peak in the sub-mm (the "sub-mm bump") and has an inverted 
($\alpha>0$) spectrum in the cm range \cite[see e.g. Nagar~\etal][]{Nag01}. 
As outflows become important, the peak 
shifts to longer wavelengths (above 1 cm), and depending on which frequencies 
were observed, the spectra might be classified as flat ($\alpha>-0.5$) 
or steep ($\alpha < -0.5$).

In this paper we report on quasi-simultaneous observations 
at 4.8, 8.6 and 19 GHz of a subset of ATESP 5~GHz sources associated
to early-type galaxies. Our results will be compared to the ones obtained 
by new high frequency ($>10$ GHz) radio surveys. 
Such surveys have started only recently, notably with 
the ATCA at 20~GHz (Ricci~\etal~\citealp{Ric04};
Sadler~\etal~\citealp{Sad06}; Massardi~\etal~\citealp{Mas08}), and detailed 
spectral studies based on high radio 
frequency observations are therefore still scarce. Other deeper 
recent high frequency surveys are
the one by Tucci \etal \cite{Tuc08}, who observed sources with $S > 20$ mJy 
at 33 GHz and compared the composition of the source counts with the high 
frequency
model counts of  De Zotti \etal \cite{Dez05}, and the 9C survey at 15 GHz 
(a 124 source sample complete to 25 mJy and a 70 source sample complete to 60 
mJy, Bolton \etal \citealp{Bol04}). In the latter case the observed 
source counts were discussed in Waldram \etal \cite{Wal07}.
The data presented in this paper represent the first 
high frequency ($>10$ GHz) systematic follow-up of a sample of 
sub-mJy radio sources.
 
The paper is organized as follows. 
We discuss the selection of the sample, the follow-up observations at 4.8, 
8.6 and 19 GHz and the data reduction in Sects.~\ref{sec:sample} and 
~\ref{sec:radiodata}; 
source variability at 4.8 GHz (where data at two epochs are available) 
is discussed in Sect.~\ref{sec:var}; a complete discussion of the radio 
spectra, in comparison with high-frequency selected samples, is given in 
Sect.~\ref{sec:results}. A further comparison with (FRI and FRII) radio 
galaxies and radio-quiet 
QSOs is presented in Sect.~\ref{sec:pdim}, where we discuss source linear 
sizes and radio powers. A brief summary of the main results of this work 
is given in Sect.~\ref{sec:sum}.
Throughout this paper we use the $\Lambda$CDM model, with 
$H_0=70$, $\Omega_m=0.3$ and $\Omega_{\Lambda}=0.7$.

\section{Sample selection} 
\label{sec:sample}

In order to better assess the composite nature of the sub-mJy 
population with particular respect to the 
low--/intermediate--luminosity AGN component, we are 
studying in great detail a sample of 131 sources with $S>0.4$ mJy 
distributed over a $2\times 0.5$ square degree region in the Southern 
sky surveyed at both 1.4 and 5 GHz in the framework of the ATESP 1.4 and 5 GHz
radio surveys \citep[Prandoni~\etal][]{Pra00a,Pra00b,Pra01a,Pra06}. 
This region overlaps entirely with the so-called 'Deep1' sub-region  
of the ESO \emph{Deep Public Survey} (DPS) survey 
(Mignano \etal \citealp{Mig08}; Olsen \etal \citealp{Ols06}).
As described in detail in Mignano \etal \cite{Mig08}, deep ($R<25$), 
multi-colour (UBVRIJK) optical/NIR data have been collected for a
part of this 
region ($1\times 0.5$ sq.~degr., namely fields Deep1a, and b) in 
the framework of the DPS and for another part ($0.5\times 0.5$ sq.~degr., 
namely field Deep1c) in the framework of the Garching-Bonn Deep Survey 
(GaBoDS, Hildebrandt \etal \citealp{Hil06}), allowing us to derive 
photometric 
redshifts and types for virtually all the radio sources optically 
identified down to $I<23.5$ in this region (Mignano \etal \citealp{Mig08}). 
Such data are complemented 
in the remaining $0.5\times 0.5$ sq.~degr. field (Deep1d) by shallower
optical 
imaging ($I<22.5$) obtained in the framework of the ESO public survey 
EIS-WIDE (Nonino \etal \citealp{Non99}) and shallower 
spectroscopic information 
($b_{j}<19.4$, Vettolani \etal \citealp{Vet98}; $I<19$, Prandoni \etal 
\citealp{Pra01b}; $I<21.5$ 
Prandoni \etal in prep). This means that in field Deep1d 
optical identification and spectral classification is available only for 
$\sim 33\%$ of the radio sources.
 
For the purposes of this work we are interested in the sources identified 
with early type galaxies. 
As reported in Table~\ref{tab:sample} (Cols.~(2)-(4)), we have 85 radio sources in fields 
Deep1a,b,c, 56 of which have been optically
identified and spectrally classified ($I<23.5$); 37 of these sources (27 
having flat/inverted spectra) are associated to early-type galaxies. 
To these sources we can add another 15 (out of 46) sources 
optically identified and spectrally 
classified in field Deep1d, ten of which are associated to early type galaxies 
(five with flat/inverted spectrum).

Subsequently we will focus our analysis on a {\it bright} sub-sample:
for sensitivity reasons we limited our multi-frequency (4.8, 8.6 and 19 GHz) 
follow-up to radio sources with $S(\rm{5 GHz})\geq 0.6$ mJy. We have a total
of 79 such sources (48 in fields Deep1a,b,c plus 31 in field Deep1d), 
41 ($32+9$) having an 
optical spectral classification and 28 ($23+5$) being classified as 
early-type (see Table~\ref{tab:sample}, Cols.~(5)-(7)). For the 6/3 cm and 12 mm follow-up
(see Table~\ref{tab:sample}, Cols.~(8)-(9)) we gave priority to the 23 sources in fields Deep1a,b,c, whose optical identification is virtually complete down to $I\sim 23.5$. 
Nevertheless multi-frequency (6/3 cm and 12 mm) 
quasi-simultaneuous observations have been undertaken also for a sub-set of 
three (out of five) Deep1d sources and for three radio sources in fields
Deep1a,b,c, which were preliminarly classified as early-type at the time of 
the 2007 observations, but not confirmed as such in the final optical 
analysis (see Mignano \etal \citealp{Mig08}). 

We notice that no target selection was done based upon the source
radio spectral shape. In other words, both flat/inverted and steep spectrum 
sources were observed. A comparison between flat, inverted and 
steep sources is important to either assess the possibly 
composite AGN component of the faint radio population, or to test for various
accretion models, from pure ADAF to ADAF+outflows/jets to pure 
jet-related accretion.

We also notice that repeating the 5 GHz observations (already available from 
the 
ATESP 5~GHz survey) allows us to test the importance of variability in the 
present sample (see Sect.~\ref{sec:var}).

\begin{table}[t]
\begin{center}
\caption{ATESP-Deep1 sample: statistics.  }
\label{tab:sample}
\begin{tabular}{c|rrr|rrr|ll}
\hline                                     
\hline
\multicolumn{1}{l}{Deep1} & \multicolumn{3}{|c}{Total Sample} &  
\multicolumn{3}{|c}{{\it Bright} Sample} & 
\multicolumn{2}{|c}{C/X,K band} \\ 
\multicolumn{1}{c}{} & \multicolumn{3}{|c}{} & \multicolumn{3}{|c}{($S\geq 
0.6$ mJy)} 
& \multicolumn{2}{|c}{Follow-up} \\
\hline
\multicolumn{1}{l}{} & \multicolumn{1}{|c}{$N_{S}$} & 
\multicolumn{1}{c}{$N_{S}^{opt}$} &  
\multicolumn{1}{c}{$N_{S}^{E}$} & \multicolumn{1}{|c}{$N_{S}$} & 
\multicolumn{1}{c}{$N_{S}^{opt}$} & \multicolumn{1}{c}{$N_{S}^{E}$} & 
\multicolumn{1}{|c}{$N_{S}^{E}$} & \multicolumn{1}{c}{$N_{S}^{\neq E \; (a)}$}\\
\hline
a,b,c & 85 & 56 & 37 & 48& 32 & 23 & 23 & 3$^{(b)}$ \\
d & 46 & 15 & 10 & 31 & 9 & 5 & $\,$ 3$^{(c)}$ & - \\
 & & & & & & & &\\
All & 131 & 71 & 47 & 79 & 41 & 28 & 26 & 3 \\ 
\hline
\multicolumn{9}{l}{\footnotesize 
$^{(a)}$ Sources for which a preliminary early-type classification has}\\ 
\multicolumn{9}{l}{\footnotesize \hspace{0.3cm}  not been confirmed}\\
\multicolumn{9}{l}{\footnotesize 
$^{(b)}$ J225048-400147; J225321-402317; J225504-400154}\\
\multicolumn{9}{l}{\footnotesize $^{(c)}$ 
J224516-401807; J224547-400324; J224654-400107}\\
\end{tabular}
\end{center}
\end{table}

\section{Radio observations and data reduction}
\label{sec:radiodata}

\subsection{Observations}
\label{sub:observations}

\begin{table}[t]
\begin{center}
\caption{Follow-up source list observed in May 2007. }
\label{tab:obs07}
\begin{tabular}{lllcc}
\hline                                     
\hline
\multicolumn{1}{c}{source name} & \multicolumn{1}{c}{R.A.} & 
\multicolumn{1}{c}{Dec} & \multicolumn{1}{c}{$t_{C/X}$} & 
\multicolumn{1}{c}{$t_{K}$} \\
     & \multicolumn{1}{c}{(J2000)}   & \multicolumn{1}{c}{(J2000)}   &  \multicolumn{1}{c}{(min)} & \multicolumn{1}{c}{(min)} \\
\hline
J225034-401936 &  22 50 34.61 & -40 19 36.3  &     5.0    &       4.0  \\
J225048-400147 &  22 50 48.04 & -40 01 47.0  &    22.5    &      10.0  \\
J225057-401522 &  22 50 57.79 & -40 15 22.5  &     7.5    &       6.0  \\
J225223-401841 &  22 52 23.82 & -40 18 41.9  &    47.5    &      42.0  \\
J225249-401256 &  22 52 49.92 & -40 12 56.0  &    35.0    &      40.0  \\
J225321-402317 &  22 53 21.28 & -40 23 17.7  &    35.0    &      62.5  \\
J225322-401931 &  22 53 22.73 & -40 19 31.6  &    50.0    &      82.0  \\
J225323-400453 &  22 53 23.89 & -40 04 53.7  &    30.0    &      10.0  \\
J225344-401928 &  22 53 44.89 & -40 19 28.6  &   167.5    &     180.0  \\
J225404-402226 &  22 54 04.33 & -40 22 26.9  &     5.0    &      14.0  \\
J225430-400334 &  22 54 30.47 & -40 03 34.0  &    50.0    &       8.0  \\
J225434-401343 &  22 54 34.70 & -40 13 43.2  &     5.0    &       4.0  \\
J225436-400531 &  22 54 36.29 & -40 05 31.3  &    67.5    &      38.0  \\
J225449-400918 &  22 54 49.73 & -40 09 18.6  &    55.0    &      66.0  \\
J225504-400154 &  22 55 04.67 & -40 01 54.8  &     5.0    &       6.0  \\
\hline
\end{tabular}
\end{center}
\end{table}

\begin{table}[t]
\begin{center}
\caption{Follow-up source list observed in July 2008. }
\label{tab:obs08}
\begin{tabular}{lllcc}
\hline                                     
\hline
\multicolumn{1}{c}{source name} & \multicolumn{1}{c}{R.A.} & 
\multicolumn{1}{c}{Dec} & \multicolumn{1}{c}{$t_{C/X}$} & 
\multicolumn{1}{c}{$t_{K}$} \\
     & \multicolumn{1}{c}{(J2000)}  & \multicolumn{1}{c}{(J2000)}  &  
\multicolumn{1}{c}{(min)} & \multicolumn{1}{c}{(min)} \\
\hline
J224516-401807 &  22 45 16.70 & -40 18 07.8  &     7.0 &    4.0   \\ 
J224547-400324 &  22 45 47.88 & -40 03 24.2  &     7.0 &    2.0   \\
J224654-400107 &  22 46 54.53 & -40 01 07.2  &     7.0 &    2.0   \\
J224750-400148 &  22 47 50.03 & -40 01 48.6  &     7.0 &    2.0   \\
J224753-400455 &  22 47 53.73 & -40 04 55.1  &   210.0 &  140.0   \\
J224822-401808 &  22 48 22.12 & -40 18 08.2  &     7.0 &    2.0   \\
J224827-402515 &  22 48 27.20 & -40 25 15.7  &    42.0 &    4.0   \\
J224919-400037 &  22 49 19.35 & -40 00 37.2  &   105.0 &   28.0   \\
J224935-400816 &  22 49 35.21 & -40 08 16.9  &    42.0 &    6.0   \\ 
J224958-395855 &  22 49 58.26 & -39 58 55.4  &    14.0 &    2.0   \\
J225004-402412 &  22 50 04.43 & -40 24 12.4  &    21.0 &    6.0   \\
J225008-400425 &  22 50 08.82 & -40 04 25.5  &    21.0 &    6.0   \\
J225239-401949 &  22 52 39.29 & -40 19 49.5  &    42.0 &   16.0   \\
J225529-401101 &  22 55 29.51 & -40 11 01.7  &    35.0 &    8.0   \\
\hline
\end{tabular}
\end{center}
\end{table}

The 4.8, 8.6 and 19~GHz observations were carried out in two separate 
campaigns in 2007 (15 sources) and 2008 (14 sources) with the ATCA. 
The expected flux densities at 8.6 and 19~GHz were computed based upon the 
source flux densities at 5~GHz 
and the spectral index between 1.4 and 5~GHz (obtained from the original 
ATESP 1.4 and 5
GHz surveys, Prandoni \etal~\citealp{Pra00b,Pra06}). 
From the expected flux densities, sensitivity and total time to be spent on 
source were derived by requesting at least a S/N = 6. 

Table~\ref{tab:obs07} shows the list of 15 sources observed in 2007 with 
target positions and observing 
times for the C/X band and the K-band observations respectively.          
The 4.8 and 8.6~GHz observations were conducted on April 30th, May 1st 
and 6th 2007. 
Both frequencies were observed simultaneously with a band-width of 
128 MHz. The ATCA operated in the 1.5C array configuration, which gives 
an angular resolution at 5 GHz comparable to the one of the original ATESP 
1.4 GHz 6-km ATCA configuration (10-15 arcsec). This is 
important since we want to derive spectral indices from source flux 
measurements. Nevertheless we include the 6-km antenna 
in the observations  in order to be able to perform a 
higher resolution analysis of the source structure.
Each source received an average of five linear cuts on a range of hour angles 
between $-4^{\rm h}$ and $+1^{\rm h}$ in order to maximize the $(u,v)$ plane 
coverage. A phase calibrator was observed every 20-25 minutes in 1-min scans.
The K-band observations were carried out on May 25th and 26th 2007 on the same list of 
sources observed at 4.8 and 8.6~GHz at two simultaneous IFs (Intermediate 
Frequencies), i.e. at 18.496 and 19.520~GHz. The band-width of both IFs was 128 MHz. 
The total observing time on source is reported in Table~\ref{tab:obs07}. 
The array 
configuration was H214: a hybrid and shorter array configuration was chosen 
in order to map the sources with a smaller number of cuts (snapshots) and 
with an angular resolution (10-15 arcsec) roughly matching the one used for the 
4.8/8.6~GHz observations. Again the 6-km antenna was included in the 
observations. 
Each target source was scanned for a total of four 
times during the two-day run in a HA range of $1^{\rm h}$ to $4^{\rm h}$. 

The list of 14 targets observed in 2008 with the source positions and 
integration times is given in Table~\ref{tab:obs08}.
The C- and X-band run was carried out in a single day on July 7th 2008 with the 
ATCA in the array configuration 1.5B. An average of seven scans was collected 
for each target source in HA range $-6^{\rm h} - +4^{\rm h}$. 
The K-band run was scheduled on July 11th and 12th 2008 in two three-hour sessions 
with the same frequency setting and configuration used in 2007. A total number 
of four cuts in the HA range $1^{\rm h} - 3^{\rm h}.5$ was collected for each 
target during the two observing sessions. The 6-km antenna was included also for the 2008 observing runs.                           
     
\subsection{Data reduction}
\label{sub:reduction}

\begin{figure}[t]
\resizebox{\hsize}{!}{\includegraphics{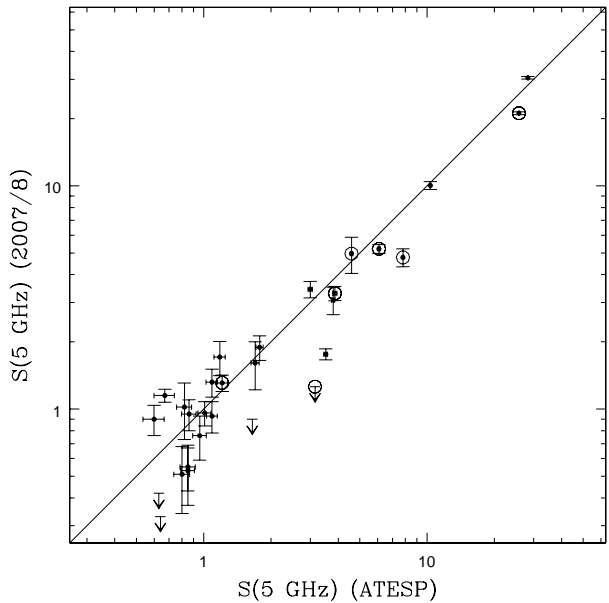}}
\caption{Comparison between ATESP 5~GHz fluxes and new 5~GHz flux measurements
obtained in 2007/2008. Upper limits are shown as downwards 
arrows. Multi-component an/or non-Gaussian extended sources are circled. The 
solid line represents $S(\rm{ATESP})=S(\rm{2007/2008})$.}
\label{fig:5GHzfluxes}
\end{figure}

\begin{table*}[t]   
\centering         
\caption{Flux densities and spectral indices of ATESP-Deep1 sample. }
\label{tab:atesp fluxes}                                         
\begin{tabular}{lrrrrrrrrrr}                       
\hline                                     
\hline                                     
Source & \multicolumn{2}{c}{ATESP Flux} & \multicolumn{6}{c}{2007/2008 Flux and rms noise values} & \multicolumn{2}{c}{spectral indices}  \\ 
            & \multicolumn{1}{c}{20 cm} &  \multicolumn{1}{c}{6 cm} & \multicolumn{2}{c}{6 cm} & \multicolumn{2}{c}{3 cm} & \multicolumn{2}{c}{1.2 cm} & \multicolumn{1}{c}{$\alpha_L$} & \multicolumn{1}{c}{$\alpha_H$} \\
 & \multicolumn{1}{c}{(mJy)} & \multicolumn{1}{c}{(mJy)} & \multicolumn{2}{c}{(mJy)} & \multicolumn{2}{c}{(mJy)} & \multicolumn{2}{c}{(mJy)} & &  \\
\hline
J224516-401807 &        9.54  &   3.87 &         3.30 &   0.24 & $<$1.44 &    0.48 & $<$0.57 &   0.19 & -0.71   $\pm$ 0.04 & $<$-1.28 $\;\;\;$$\;\;\;$$\;\;\;$$\;\;\,$ \\
J224547-400324 &       32.83 &            28.28 &                 30.49 &            0.42 &     26.64 &             0.60 &     27.89 &            0.84 & -0.12  $\pm$ 0.01 & 0.06  $\pm$ 0.01\\
J224654-400107 &         5.59 &              3.15 &              $<$1.26 &            1.00 & $<$1.44 &             0.48 &       1.11 &            0.26 & -0.45  $\pm$ 0.05 & -- $\;\;\;\;\;$$\;\;\;$$\;\;\;$$\;\;\,$  \\
J224750-400148 &       13.44 &               6.09 &                   5.22 &            0.23 &       2.63 &             0.17 &       0.98 &            0.30 & -0.62  $\pm$ 0.02 & -1.25  $\pm$ 0.29 \\
J224753-400455 &         2.08 &              0.67 &                   1.15 &            0.08 &       0.43 &             0.11 &       0.27 &            0.06 & -0.89  $\pm$ 0.19 & -0.59  $\pm$ 0.34 \\
J224822-401808 &       19.08 &            10.34 &                10.04 &            0.42 &       7.29 &             0.61 &       4.24 &            0.31 & -0.48  $\pm$ 0.01 & -0.69  $\pm$  0.04 \\
J224827-402515 &         0.58 &               0.80 &                    0.51 &            0.17 &       1.12 &             0.25 & $<$0.78 &            0.26 & 0.25  $\pm$ 0.27 & $<$-0.46  $\;\;\;$$\;\;\;$$\;\;\;$$\;\;\,$\\
J224919-400037 &         0.91 &               0.64 &              $<$0.33 &            0.11 &       0.69 &             0.16 &       0.43 &            0.10 & -0.28  $\pm$ 0.23 & -0.60  $\pm$  0.32 \\
J224935-400816 &         0.70 &               0.82 &                   1.02 &            0.29 & $<$0.57 &             0.19 & $<$0.60 &            0.20 & 0.12  $\pm$ 0.23 & $<$-0.39 $\;\;\;$$\;\;\;$$\;\;\;$$\;\;\,$ \\
J224958-395855 &         1.52 &              1.65 &              $<$0.90 &            0.30 & $<$1.02 &             0.34 &       1.52 &            0.34 & 0.06  $\pm$ 0.12 & -- $\;\;\;\;\;$$\;\;\;$$\;\;\;$$\;\;\,$ \\
J225004-402412 &         3.16 &              1.78 &                   1.89 &            0.24 & $<$0.84 &             0.28 &       1.28 &            0.21 & -0.45  $\pm$ 0.08 & -0.28  $\pm$ 0.07 \\
J225008-400425 &         2.88 &              1.70 &                   1.61 &            0.39 & $<$0.84 &             0.28 & $<$0.66 &            0.22 & -0.41  $\pm$ 0.09 & $<$-0.65 $\;\;\;$$\;\;\;$$\;\;\;$$\;\;\,$ \\
J225034-401936 &       76.62 &            25.78 &                21.16 &            0.30 &     17.20 &             0.85 &       6.03 &            0.33 & -0.86  $\pm$ 0.01 & -1.33  $\pm$ 0.02 \\
J225048-400147$(^a)$ &  0.60 &  0.96 &  0.76 & 0.17 &  1.13  &  0.23 & $<$0.51    & 0.17 & 0.37  $\pm$ 0.27 & $<$-1.01 $\;\;\;$$\;\;\;$$\;\;\;$$\;\;\,$ \\
J225057-401522 &         2.01 &            3.00 &                 3.44 &            0.29 &       2.16 &             0.40 &       2.51 &            0.28 & 0.31  $\pm$ 0.08 & 0.19  $\pm$ 0.14 \\ 
J225223-401841 &         0.98 &              0.85 &                   0.55 &            0.12 & $<$0.51 &             0.17 &       0.46 &            0.13 & -0.11  $\pm$ 0.20 & -0.13  $\pm$ 0.21 \\ 
J225239-401949 &         2.26 &             1.18 &                   1.71 &            0.30 &       1.02 &             0.25 & $<$0.42 &            0.14 & -0.51  $\pm$ 0.12 & $<$-1.13 $\;\;\;$$\;\;\;$$\;\;\;$$\;\;\,$ \\
J225249-401256 &         1.52 &               1.09 &                   0.93 &            0.15 &       0.84 &             0.21 &       0.71 &            0.10 & -0.26  $\pm$ 0.14 & -0.21  $\pm$ 0.24 \\
J225321-402317$(^a)$  &  2.32 &  1.21 &  1.31 &  0.11 & $<$0.57 &  0.19 & 0.31  &    0.09 & -0.51  $\pm$ 0.11 & -1.05  $\pm$ 0.15 \\ 
J225322-401931 &         1.86 &              1.01 &                   0.96 &            0.12 &       0.92 &             0.17 &       0.43 &            0.09 & -0.48  $\pm$ 0.14 & -1.06  $\pm$ 0.23 \\ 
J225323-400453 &         0.51 &              0.85 &                   0.53 &            0.16 & $<$0.69 &             0.23 &       0.75 &            0.23 & 0.40  $\pm$ 0.31 & 0.25  $\pm$ 0.31 \\ 
J225344-401928 &         0.60 &              3.52 &                   1.76 &            0.10 &       2.12 &             0.14 &       2.05 &            0.08 & 1.39  $\pm$ 0.25 & -0.04  $\pm$ 0.02 \\  
J225404-402226 &       10.34 &              3.80 &                   3.08 &            0.43 &       3.79 &             0.64 &       0.78 &            0.21 & -0.79  $\pm$ 0.03 & -2.01  $\pm$ 0.30 \\ 
J225430-400334 &   $<$0.26 &              0.63 &             $<$0.42 &            0.14 &       1.43 &             0.20 &       1.26 &            0.29 & $>$0.70 $\;\;\;$$\;\;\;$$\;\;\;$$\;\;\,$  & -0.16  $\pm$  0.21 \\ 
J225434-401343 &       21.09 &               7.80 &                   4.78 &            0.44 &       2.95 &             0.64 &       2.67 &            0.43 & -0.78  $\pm$ 0.01 & -0.13  $\pm$ 0.21 \\ 
J225436-400531 &         0.47 &             0.60 &                  0.90 &            0.14 &       0.54 &             0.15 & $<$0.45 &            0.15 & 0.19  $\pm$ 0.36 & $<$-0.23 $\;\;\;$$\;\;\;$$\;\;\;$$\;\;\,$ \\ 
J225449-400918 &         1.24 &             0.86 &                   0.95 &            0.15 &       0.53 &             0.16 & $<$0.36 &            0.12 & -0.29  $\pm$ 0.17 & $<$-0.49 $\;\;\;$$\;\;\;$$\;\;\;$$\;\;\,$ \\ 
J225504-400154$(^a)$  &   9.67 &   4.59 &  4.97 & 0.92 & 4.91 & 1.53 & 
$<$1.17  & 0.39 & -0.59  $\pm$ 0.03 & $<$-1.82 $\;\;\;$$\;\;\;$$\;\;\;$$\;\;\,$ \\
J225529-401101 &         1.48 &              1.09 &                 1.32 &            0.19 &       1.34 &             0.27 & $<$0.63 &            0.21 & -0.24  $\pm$ 0.14 & $<$-0.96 $\;\;\;$$\;\;\;$$\;\;\;$$\;\;\,$ \\ 
\hline
\multicolumn{10}{l}{$(^a)$ Sources not identified with early-type galaxies.}
\end{tabular}                                                
\end{table*} 

The raw visibility data were reduced using the astronomical software package 
MIRIAD (Sault \etal 1995). The data were read into MIRIAD, checked, 
flagged and split into single-source single-frequency files.
The bandpass was equalised using the standard ATCA calibrator 1921-293 (apart 
from 2007 K-band run where 2243-123 was used instead). 
The absolute flux scale was bootstrapped using 
the standard ATCA primary 
calibrator 1934-638. The phase calibrators 2226-411 and 2232-488 (only for the 
2008 K-band run) were used 
to solve for the amplitude and phase complex gains in each observing 
frequency. 
The calibrated visibilities 
were then imaged by Fourier inversion, {\it after excluding the 6-km antenna}: 
the 4.8 and 8.6~GHz visibilities were imaged separately. The K-band
visibilities were instead combined into a single 19-GHz image by using the 
Multi Frequency Synthesis (MFS)
technique (Sault \& Wieringa \citealp{Sau94}). 
In the case that the observing frequencies of two or 
more IFs were not too far 
apart so that radio spectral shape effects were still negligible, the MFS 
technique could be used effectively 
to increase the $(u,v)$ plane coverage and minimise the band-width smearing 
effects and thus improve the 
final image quality. As we were carring out a detection experiment, inverted
images were inspected using the 
KARMA package visualisation tool {\it kvis} to search for at least a 3-sigma 
detection in the inverted image 
position were the source was expected to be found (the field centre). 
In case of detection the source images 
were cleaned and restored using standard clean algorithms: H{\"o}gbom 
\cite{Hog74}, 
Clark \cite{Cla80} or Steer (Steer, Dewdney, \& Ito \citealp{Ste84}). The MFS 
technique was used only at the Fourier inversion stage, 
{\it not} at the cleaning stage.

Limited $(u,v)$ angle coverage of the C/X band observations
produced elongated 
restoring beams ($24''\times 6''$ at 4.8 GHz and $14''\times 3.5''$ at 8.6 
GHz in 2007;
$18''\times 7''$ at 4.8 GHz and $10''\times 4''$ at 8.6 GHz in 2008), 
making the cleaning unreliable. We therefore 
decided to determine the source fluxes by directly fitting a Gaussian 
point-like 
model to the calibrated visibilities. In order to check for the presence of 
extended structure, we fitted the visibilities at different 
angular resolutions. We compared the fluxes obtained by either including or 
removing 
the 6-km antenna, or, for the 8.6 GHz observations only, by removing all 
baselines 
longer than 750 m (corresponding to 25 k$\lambda$) in order to have a similar 
angular 
resolution as for the 4.8 GHz observations ($20''\times 6''$). 

From this analysis we found that most of the sources are detected 
($S/N\geq 3$) at either 
4.8 or 8.6 GHz or both, and that most of 
the detected sources can be 
reliably considered point-like at the lowest angular resolution 
(no 6-km antenna at 4.8 GHz and no baselines longer than 750 m at 8.6 GHz). 
For those sources (peak) flux densities were determined by fitting a 
point-like Gaussian 
to the lowest resolution datasets. The remaining sources (typically 
well-resolved or 
multi-component in the original ATESP 5 GHz catalogue,  Prandoni \etal 
\citealp{Pra06}) were 
uv-fitted with a single Gaussian whenever suitable, or were alternatively fitted 
with 
multi-component Gaussians in the restored images (convolved to the same 
resolution as 
for the other frequencies). 
For such resolved sources integrated flux densities were considered to derive 
the 
spectral index. 

The $(u,v)$ plane coverage obtained for the K-band observations is much better,
thanks to 
the availability of North-South baselines of the hybrid ATCA configuration 
(restoring 
beam $\sim 10''\times 8''$, excluding the 6-km antenna). Therefore
K-band flux densities were derived by performing source Gaussian fitting 
in the 
image plane. Peak (or integrated) flux densities were used for unresolved 
(or resolved) 
sources to derive the spectral index.

For all undetected sources $3\sigma$ upper limits were derived by estimating 
the noise 
from V Stokes parameter visibilities for the C/X band observations and 
directly from the 
images for K-band observations. 

In Table \ref{tab:atesp fluxes} we summarise the flux densities at the 
available four 
frequencies for all the 29 sources observed. The three sources not 
identified with early type galaxies (see discussion in Sect.~\ref{sec:sample})
are explicitely indicated in the Table. 
For the new observations we list also the rms noise values estimated 
directly from the images (for K-band observations) 
or from V Stokes 
parameter visibilities (for C/X observations). ATESP observations were 
designed to have very uniform image rms values of the order of $0.08-0.09$ 
mJy at 1.4 GHz
and $0.06-0.08$ mJy at 5 GHz (Prandoni \etal \citealp{Pra00a,Pra06}).
The rms noise values can be 
used to estimate source signal-to-noise ratios and uncertainties 
in the flux density measurements.  
Table \ref{tab:atesp fluxes} also lists $\alpha_L$ and $\alpha_H$, 
the spectral indices derived at low and high frequency respectively.
In particular we used the original ATESP 1.4 and 5~GHz flux densities to 
derive $\alpha_L$ and the new simultaneous    
8.6 and 19 GHz measurements to derive $\alpha_H$.
We notice however that for sources undetected at 8.6~GHz,
$\alpha_H$ was derived by using the simultaneous 4.8 GHz flux measurement, if 
available.
Radio spectra and contour plots for the 26 sources associated to early-type 
galaxies are presented at the end of the Paper in Figs.~\ref{fig:fluxpanel}
and \ref{fig:overlays}. 

Figure~\ref{fig:5GHzfluxes} compares the ATESP 5~GHz fluxes (Prandoni \etal 
2006) with the 5~GHz fluxes obtained from the multi-frequency observation 
campaign presented here. The two epoch fluxes agree reasonably well
with larger scatters for fainter sources (generally corresponding to lower 
signal-to-noise ratios). The highest deviations from the expected $1:1$ 
line at high flux levels are found for multiple/extended sources (circled dots)
where new integrated flux determinations, based on poor $(u,v)-$angle 
coverage data, may be underestimated due to the presence of undetected low 
surface brightness extended flux. In Sect.~\ref{sec:var} we will also 
explore the possible role of source variablity in producing significant 5~GHz 
flux deviations in single-component sources.

\subsection{Full resolution K-band imaging}
\label{sub:kband}

\begin{figure*}
\includegraphics[scale=0.8]{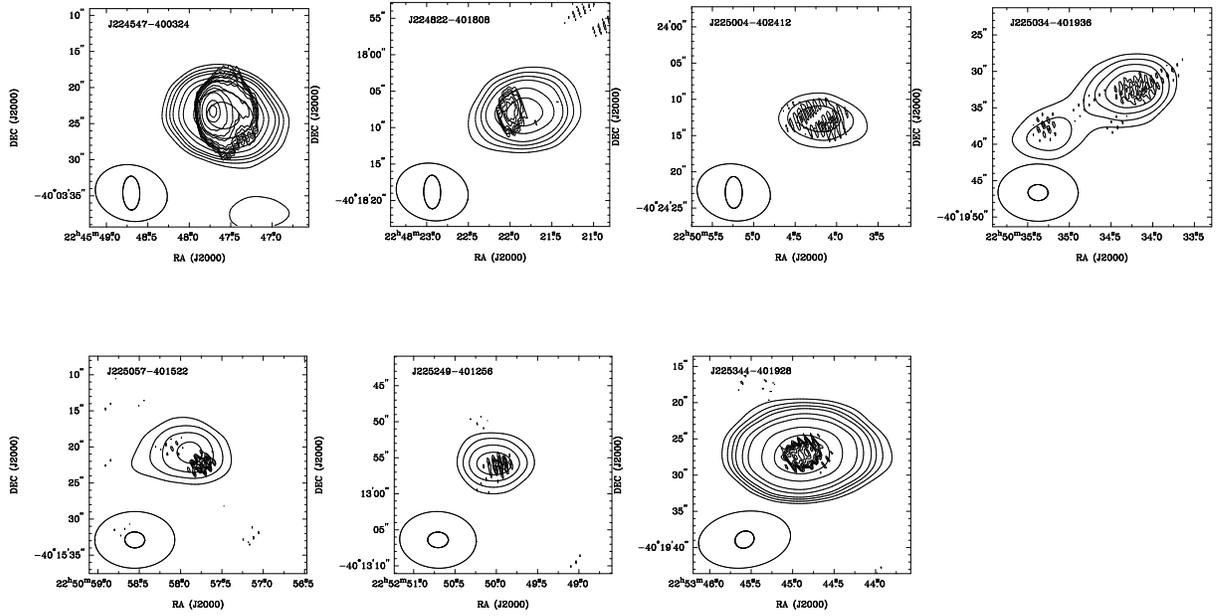}

\vspace{-12.0cm}
\caption{K-band full resolution contours (thick lines) superimposed on 
low resolution contours (thin lines) for the seven sources detected in 
low resolution K-band images with $S/N>6$.
Contour levels are 3, 4, 5, 6, 8, 10, 15, 20, 30, 50 and 60 \% of 
the peak flux density.}
\label{fig:fullres}
\end{figure*}

ATESP 5~GHz sources were originally imaged at both low ($\sim 10''$)
and full ($\sim 2''$) resolution \cite[Prandoni~\etal~][]{Pra06}, 
in order to study their structure (see contours in Fig.~\ref{fig:overlays}). 
Here we exploit the K-band data to repeat this kind of 
analysis at a higher frequency. K-band images were produced at full 
resolution ($\sim 2''$)
by including the 6-km antenna in the imaging and cleaning process. 
This has been done for all sources detected in K-band low resolution images 
with $S/N>6$ (7 in total). 
Results are reported in Table~\ref{tab:fullres} and Fig.~\ref{fig:fullres}, 
where K-band contour plots at 
low (thin contours) and full (thick contours) resolution are shown. 
Unfortunately the full resolution K-band imaging does not provide clear 
morphological information for any source, except for the source J224547-400324, which
shows an elongated structure in east-west direction. 

\begin{table}[t]
\begin{center}
\caption{Full resolution K-band imaging. Source parameters. }
\label{tab:fullres}
\begin{tabular}{lllrr}
\hline                                     
\hline
\multicolumn{1}{c}{source name} & \multicolumn{1}{c}{R.A.} & 
\multicolumn{1}{c}{Dec} & \multicolumn{2}{c}{$S_{int}\pm 1\sigma$} \\
     & \multicolumn{1}{c}{(J2000)} & \multicolumn{1}{c}{(J2000)}   
&  \multicolumn{2}{c}{(mJy)} \\
\hline
J224547-400324 &  22 45 47.67 & -40 03 23.4  &     25.70 &  $\pm 0.22$   \\
J224822-401808$(^a)$ &  22 48 21.99 & -40 18 07.8  &  3.22 &   $\pm 0.24$  \\
J225004-402412$(^b)$ &    &  &    1.70 &    $\pm 0.20$   \\
J225034-401936$(^b)$ &    &  &    7.15 &   $\pm 0.31$   \\
J225057-401522 &  22 50 57.73  & -40 15 22.3 &    2.21 &   $\pm 0.24$   \\
J225249-401256$(^b)$ &    &  &    0.63 &   $\pm 0.13$   \\
J225344-401928 &  22 53 44.89  & -40 19 27.2 &    2.17 &   $\pm 0.08$   \\
\hline
\multicolumn{5}{l}{\scriptsize{$(^a)$ Low surface brightness flux present to the right of the source. Including it we get}} \\ 
\multicolumn{5}{l}{\hspace{0.3cm} \scriptsize{$S=4.3$ mJy.}} \\
\multicolumn{5}{l}{\scriptsize{$(^b)$ Coordinates are given only for Gaussian fitted sources. Fluxes for multi-component  }} \\ 
\multicolumn{5}{l}{\hspace{0.3cm} \scriptsize{and/or low $S/N$ sources were obtained by summing the pixels over the entire source.}}\\
\end{tabular}
\end{center}
\end{table}

\begin{figure}[t]
\resizebox{\hsize}{!}{\includegraphics{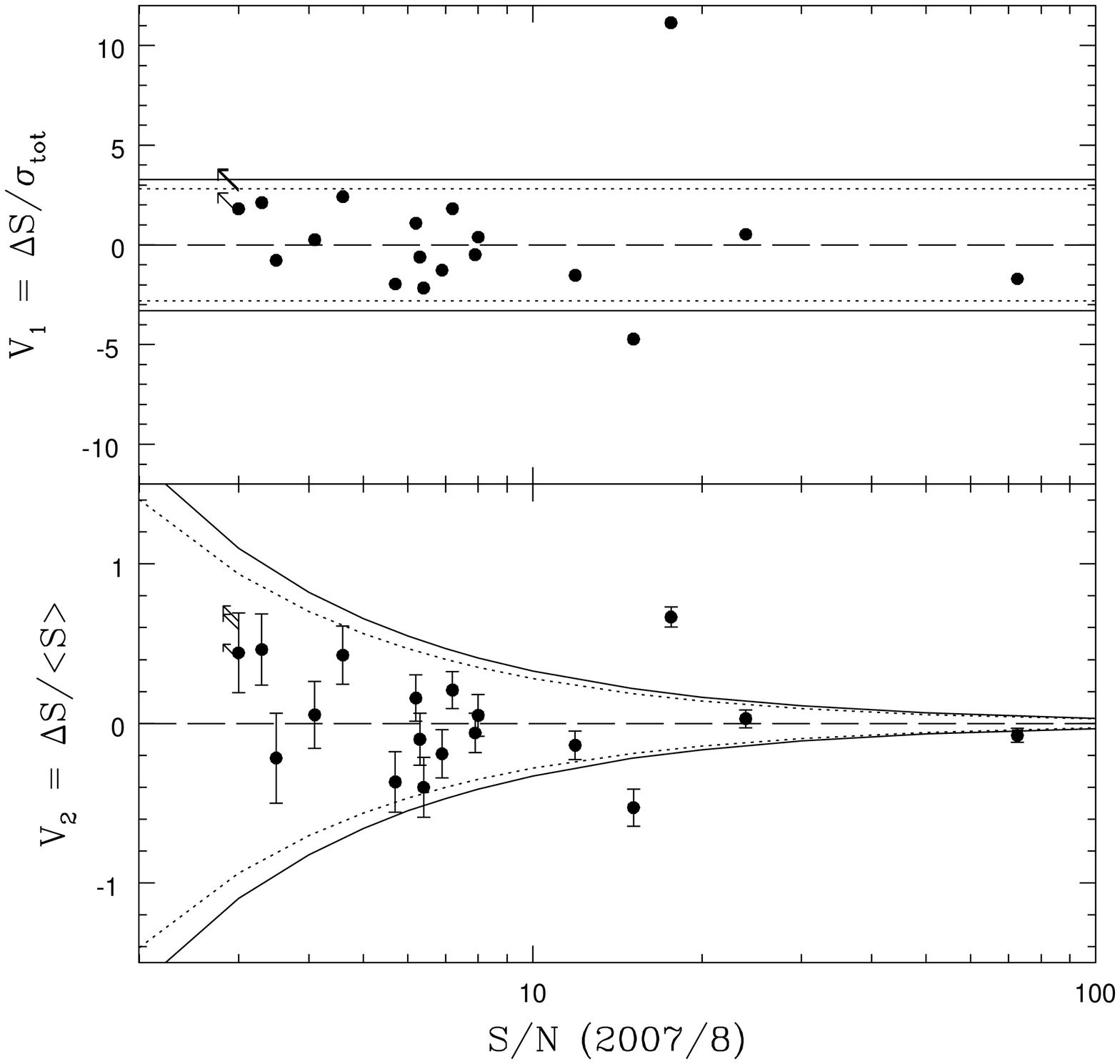}}
\caption{Variability index $V_1$ (Top) and $V_2$ (Bottom) for single-component 
sources as a function of S/N (referred to 2007/8 
observations). Arrows indicate sources which were not detected in epoch 2. 
Dotted and solid lines correspond to random $\Delta S$ fluctuation 
probabilities of 0.5\% and 0.1\% respectively.
\label{fig:var}}
\end{figure}

\section{Source variability}
\label{sec:var}

Before proceeding with the analysis of the radio spectral properties of 
our sample we should consider the possible effect of a source variability in 
shaping our radio spectra (presented later in Fig.~\ref{fig:fluxpanel}). 
Although high frequency 
observations were done simultaneously in either 2007 or 2008, low 
frequency observations were done long before that, between 1999 and 2005. 

From Sadler \etal 
\cite{Sad06} we know that high frequency (K-band) variability is not very 
important on time scales of a few years: only 5 ~\% of the 
sources turn out to be variable at a level $> 30$~ \% over a baseline of 1-2 
years. We assume that at the lower frequencies (1.4 and 4.8 GHz) the 
variability will be lower than that, since the low frequency emission may 
mostly come from extended, optically thin regions. 
However, we cannot exclude that in ten 
years time the overall level of emission at high frequencies has changed enough 
so that, for example, the spectral index 
$\alpha_M$ (between 5 and 8.6~GHz) may have varied while $\alpha_L$ (between 
1.4 and 5~GHz) and $\alpha_H$ (between 8.6 and 19~GHz) remained (roughly) 
constant. Such a change may manifest itself if, 
for example, both the high and low frequency spectral indices are negative, 
but the intermediate spectral index is positive. 
This kind of effect does indeed happen, as a visual inspection of 
Fig.~\ref{fig:fluxpanel} easily reveals (see e.g. sources J225057-401522 and 
J225404-402226),
although it is not so big as to influence the discussion in a significant way. 
This confirms and strengthens the 
conclusion of Sadler \etal \cite{Sad06} that 
strong source variability is a minor effect, even on time scales of the order 
of 5-10 years. 

For a more quantitative analysis of the source variability in our sample, 
we compared 
the original ATESP 5 GHz 
fluxes (obtained in 2000/2001, see Prandoni \etal~\citealp{Pra06}) 
with the ones obtained with the 2007/2008 follow-up and checked
the significance of their variations through the analysis of the so-called 
{\it variability index}. In particular we analysed two 
well-known variability indices, defined as follows (see 
Gregorini \etal \citealp{Gre86}; Quirrenbach \etal \citealp{Qui92}):

\begin{eqnarray*}
V_1 &=& \Delta S/\sigma_{tot} \\
V_2 &=& \Delta S/<S>  
\end{eqnarray*}

\noindent
where $\Delta S = S_1-S_2$, with $S_1$ and $S_2$ being the 5 GHz fluxes 
measured respectively in epochs $1$ (2000/01) and $2$ (2007/08);   
$\sigma_{tot}=\sqrt{\sigma^2(S_1)+\sigma^2(S_2)}$, and
$<S>= \frac{S_1+S_2}{2}$. We notice that $V_2$ represents the 
so-called {\it modulation index} as it reduces in the case of two epochs measurements. 

Figure~\ref{fig:var} shows the $V_1$ and $V_2$ variability index 
distribution as a function of the signal-to-noise ratio for 
single-component sources (multi-component sources are unlikely to vary). 
To compute $V_1$ we estimated $\sigma(S_1)$ and $\sigma(S_2)$ through 
Condon's 
master equations (Condon~\citealp{Con97}) by assuming 
calibration errors of the order of 3\% for both epochs (see 
Prandoni \etal~\citealp{Pra00b}). 
Dotted and solid lines in Fig.~\ref{fig:var} correspond to random 
Gaussian $\Delta S$ fluctuation probabilities of 0.5\% and 0.1\% respectively.  
In general a source is considered as a good {\it variable} candidate whenever
the above probability is $<0.1\%$ (or $\Delta S> 3.29\sigma_{tot}$). It is 
clear from the figure that two sources (J224753-400455 and J225344-401928)
satisfy both the $V_1$ and $V_2$ variability criteria, while a third source 
(J224547-400324) is defined as {\it variable} only with respect to $V_2$ and 
for this reason has to be considered a less reliable candidate. 
All the other sources have 
5~GHz flux variations consistent with being random fluctuations.

From the analysis of the overall source spectra 
(see Fig.~\ref{fig:fluxpanel}) we see 
that source J224753-400455 ($\Delta S=-4.7\sigma_{tot}$; $\Delta S= -0.53<S>$) 
has a smooth power-law spectrum from 1.4 to 19 GHz, with the only exception
of the 5~GHz point obtained in 2008. This argues against variability for this 
sources. On the other hand source J225344-401928 ($\Delta S=11.1\sigma_{tot}$;
$\Delta S= 0.67<S>$), clearly shows a change in the spectral shape from 
strongly inverted to flat when we go from the ATESP
non-simultaneous flux measurements (red squares) to the present 
quasi-simultaneous flux determinations (blue diamonds). This strongly 
supports the hypothesis of variability. 
Source J224547-400324 is a less reliable candidate 
($\Delta S=-1.7\sigma_{tot}$) and flux variations, if real, are quite small 
($\Delta S= -0.08<S>$). 
The overall spectrum is rather flat over the entire 
frequency range covered. Both the flatness of the
spectrum and the radio morphology (not inconsistent with a core-jet structure,
see Fig.~\ref{fig:fullres}) may support the presence of some level of
variability in this source.

In summary, we find that one of the 21 (e.g. 5\%) single-component sources 
in our sample is a strong variable, while a second source may present some 
low-level variability. 
This is consistent with previous results. Tingay et al~\citep{Tin03} found
that the level of variability of powerful ($\sim 1$ Jy), compact radio sources 
moderately 
increases with frequency from a median variability of 6\% at 1.4 GHz to 9\%
at 8.6 GHz over a time-scale of $3-4$ years. Barvainis \etal~\citep{Bar05} 
found similar variability levels at 8.5 GHz for a sample of radio-loud and
radio-quiet QSOs observed with the Very Large Array. It is worth noting that 
the flux density levels probed by the radio-quiet QSOs of the Barvainis \etal
sample are similar to the flux densities of our sources ($\lsimeq 1$ mJy).

\section{Discussion of radio spectra}
\label{sec:results}

\begin{figure*}[t]
\includegraphics[scale=0.4]{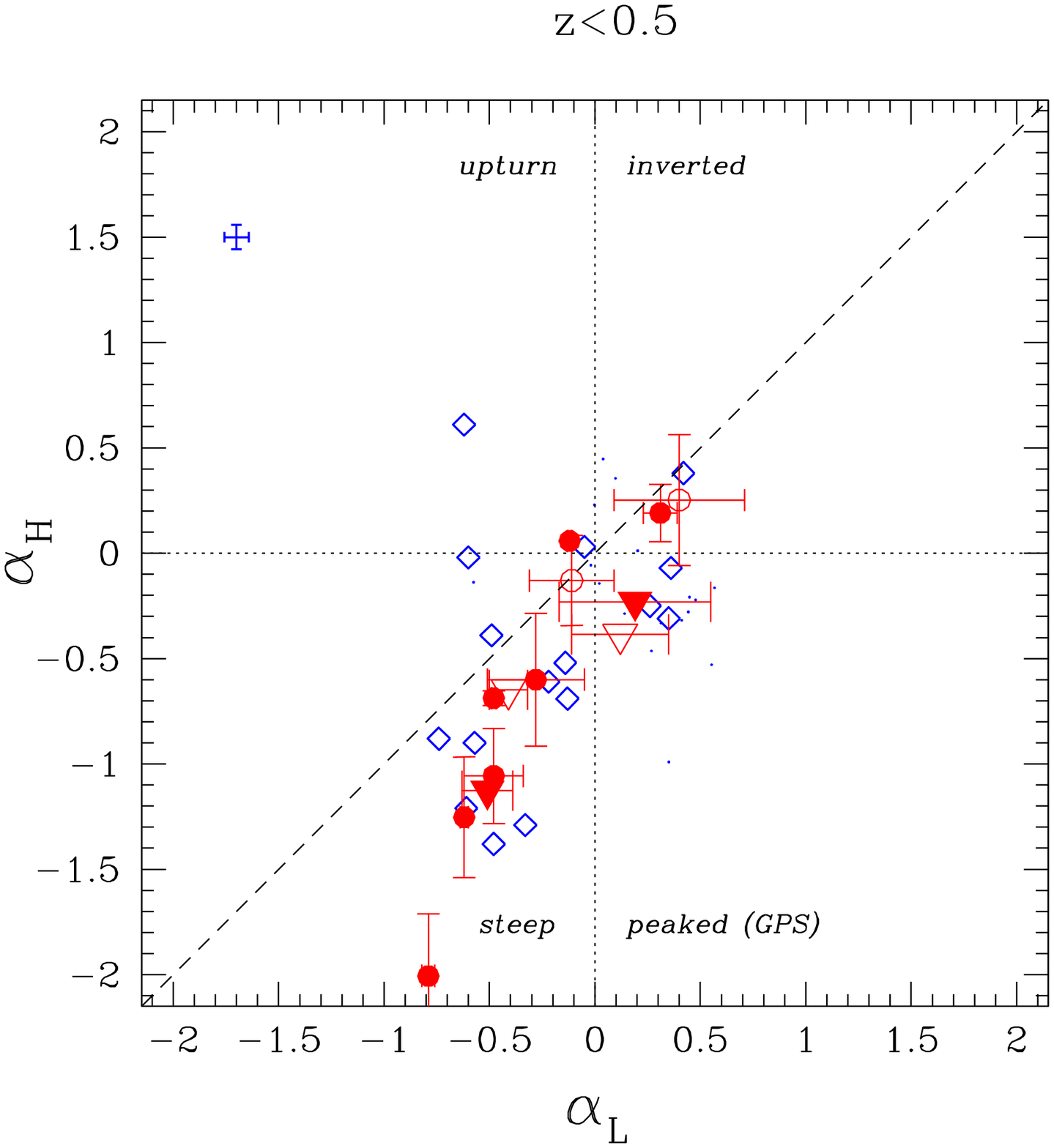}\includegraphics[scale=0.4]{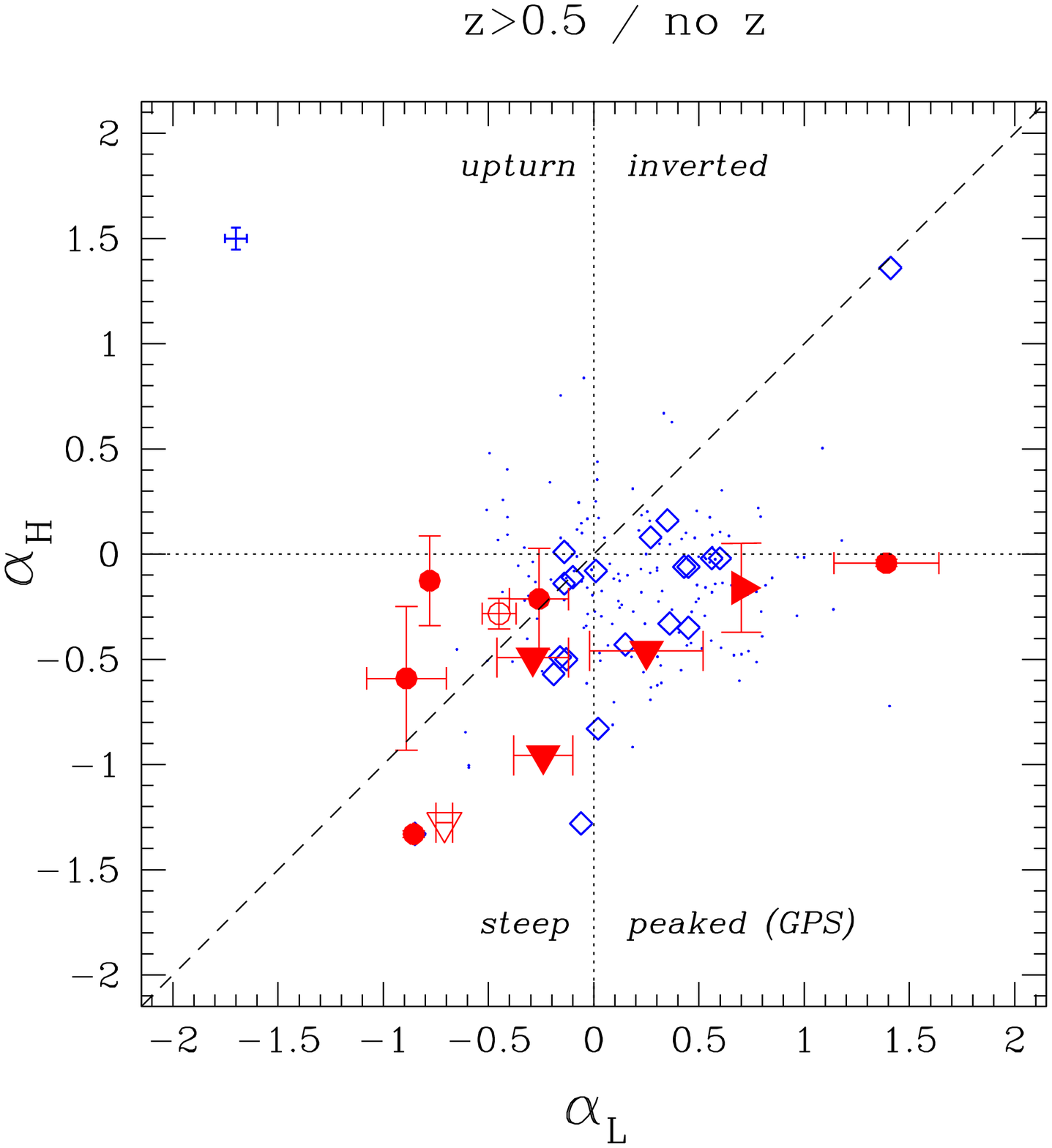}

\caption{
High-- against low--frequency spectral index. 
Red symbols indicate ATESP sources: filled symbols are used for 
$\alpha_H$ derived between 8.6 and 19 GHz and empty symbols are used 
for $\alpha_H$ derived between 4.8 and 19 GHz; triangles indicate upper/lower 
limits in $\alpha_H/\alpha_L$ respectively. Blue diamonds indicate sources 
associated with early type optical spectra from  the Massardi \etal \cite{Mas08} 
sample; blue dots indicate sources associated with quasars or quasar 
candidates in the Massardi \etal \cite{Mas08} sample. The typical error box for
the Massardi \etal sources is indicated in the upper left corner.
{\it Left panel:} 
objects with redshift $< 0.5$. {\it Right panel:} objects with either 
redshift $>0.5$, or unknown redshift. 
The dashed line represents the location of sources with power-law 
spectra (i.e. with $\alpha_H=\alpha_L$). 
Horizontal and vertical dotted lines respectively 
indicate the $\alpha_H= 0$ and $\alpha_L= 0$ lines defining the four 
quadrants discussed in the text.}
\label{fig:alfapanel}
\end{figure*}

Figure~\ref{fig:fluxpanel} shows the overall radio spectra of the 26 ATESP 
5~GHz sources associated with early-type galaxies discussed in this work. 
Red circles represent non-simultaneous ATESP 1.4 and 5 GHz observations, while
blue diamonds represent the new quasi-simultaneous observations
at 4.8, 8.6 and 19 GHz. We notice that
low frequency spectral behaviour (steep or flat) derived from 1.4 and 5 GHz 
ATESP data is generally confirmed by high frequency simultaneous 
observations. 
This can be considered as a first qualitative confirmation of the fact that the 
sub-mJy flat-/inverted-spectrum population associated with early-type 
galaxies revealed by the ATESP 1.4 and 5~GHz surveys 
(Prandoni~\etal\citealp{Pra06}; Mignano~\etal\citealp{Mig08}) and discussed in 
Sect.~\ref{sec:introduction} is genuine. The most notable exception 
is represented by the {\it variable} source J225344-401928 (see 
Sect.~\ref{sec:var}),
which is remarkably characterised by a very inverted spectrum between 1.4 and 
5 GHz ($\alpha_L=1.4$), not confirmed as real by high frequency simultaneous 
observations ($\alpha_M\sim \alpha_H\sim 0$).    

For a more detailed analysis of the radio spectra we will focus below
on two spectral indices, $\alpha_L$ and $\alpha_H$ (listed
in Table~\ref{tab:atesp fluxes}). These two spectral indices 
can be used in a sort of "two-colour" 
diagram as was done in previuos works (see Sadler \etal \citealp{Sad06}; 
Massardi et al \citealp{Mas08}; Tucci \etal \citealp{Tuc08}). 
Note that the analysis of the two--colour diagram is not exhaustive, and 
the spectral index between 4.8 and 8.6 GHz (e.g. $\alpha_M$) may give further 
independent information on the spectrum, as previously noted in 
Sect.~\ref{sec:var}.

We will also compare our data with recent high frequency 
surveys (Sadler \etal \citealp{Sad06}; 
Massardi \etal \citealp{Mas08}; Tucci \etal \citealp{Tuc08}) and will try to 
determine how and to what extent frequency selection and flux limits 
influence the results. 
The ATESP 5~GHz survey is the only one selected at low frequency, 
while the other 
three are based on $20-30$ GHz observations. 
Also there is quite a difference in the flux limit, 
which reaches from $\sim 0.6$ mJy (our sample) via 20 mJy at 33 GHz (Tucci et 
al. \citealp{Tuc08}), 100 mJy at 20 GHz (Sadler \etal \citealp{Sad06}) to 
500 mJy at 20 GHz (Massardi \etal~\citealp{Mas08}). 
One extra selection criterion used for our sample was that we followed-up at 
high frequencies only sources associated with galaxies having early-type 
spectra; for that reason we also defined a sub-sample of 36 sources associated 
with galaxies from the Bright Source Sample
discussed by Massardi \etal \cite{Mas08}. Since the latter sources are 
all bright 
in the radio, these are presumably all classical radio galaxies; their typical 
radio powers put most of them effectively in the FRII range.  

Figure \ref{fig:alfapanel} shows two--colour diagrams of our (red symbols) and 
the Massardi~\etal galaxy (blue diamonds) samples. Also plotted are all sources 
from the Massardi~\etal sample that are associatd with quasars or
quasar candidates (blue dots). 
We divided the sources in a 
nearby ($z <0.5$, left panel) and 
a far ($z > 0.5$, right panel) sample, in order to check if distance may 
have an effect 
upon the spectral indices. In the far sample we included quasar candidates with
unknown redshift, the large majority of which (presumably) are at 
redshifts $>0.5$ on the basis of their optical magnitude distribution. 

For a physical interpretation of the two-colour diagram we divided it in four 
quadrants, as done in previous works: 1) sources 
with $\alpha_L<0, \; \alpha_H<0$, 
conventionally named {\it steep}\footnote{In fact in this quadrant fall both
steep ($\alpha<-0.5$) and flat ($-0.5<\alpha<0$) sources}; 
2) sources with $\alpha_L<0, \; \alpha_H>0$, named {\it upturn}; 
3) sources with $\alpha_L>0, \; \alpha_H>0$, named {\it inverted}; 
4) sources with $\alpha_L>0, \; \alpha_H<0$, named {\it peaked}. 
We point out that the radio spectral index of the overall 
synchrotron emission from a jet-dominated radio source is expected 
to be from steep to moderately inverted ($-0.7<\alpha<0.2$) at the frequencies 
of this work, depending on the 
relative contributions of extended (optically-thin) and base (self-absorbed) 
jet components. Jet-dominated sources are therefore expected to mostly 
fall into the {\it steep} quadrant. Instead, for an ADAF, one 
expects $0.2 < \alpha < 1.1$ up to mm wavelengths 
(see Nagar~\etal\citealp{Nag01}), with $\alpha$ varying 
according to the accretion rate (f.i. $\alpha = 0.4$ 
when $L\sim 10^{-4}L_{\rm{Edd}}$ and $\alpha \sim 1$ when 
$L\leq 10^{-7}L_{\rm{Edd}}$; Mahadevan~\citealp{Mah97}). ADAF systems are 
therefore expected to fall into the {\it inverted} quadrant. Co-existence of 
moderate outflows can flatten the spectrum (with $\alpha$ remaining positive), 
whereas strong outflows can shift the peak of the radio emission from mm to cm 
wavelengths (Quataert \& Narayan~\citealp{Qua99}). 

As can be seen in Fig.~\ref{fig:alfapanel} our ATESP sources, as well as the 
sources from the
much brighter Massardi~\etal galaxy sample, 
are mostly located in  the lower part of the plots: 23/24 and 33/36 sources 
for the ATESP and the Massardi sample respectively have $\alpha_H <0.2$. 
Although these numbers are significant only at about the $3\sigma$ level 
due to the relatively poor statistics involved, there can be little 
doubt that both samples consistently contain more sources with 
$\alpha_H < 0.2$. This is of course hardly surprising
for the ATESP sample, which was originally selected at 5~GHz, but the 
high frequency selection of the Massardi
sample does not make much difference, at least in this respect.
Less obvious a priori is the general steepening ($\alpha_L>\alpha_H$) of the 
spectra from low to high frequency. 
This is evident from the large number of sources lying below the diagonal 
dashed line, 
corresponding to power-law spectra (19/24 and 31/36 for ATESP and Massardi). 
These numbers are significant only at the $2\sigma$ level, but again both 
samples behave consistently and 
actually appear to cover identical regions in the two-colour diagram.
Both facts
favour a jet-dominated scenario, with steep-spectrum 
($\alpha_H<-0.5$) sources dominated by synchrotron optically-thin emission 
and flat-spectrum ($\alpha_H>-0.5$) sources dominated by synchrotron 
optically-thick emission 
coming from the base of the jet. Nevertheless jet+ADAF 
models can also be consistent with values 
$0\lsimeq\alpha_H \lsimeq 0.2$. Pure ADAF models 
($\alpha_{L}$ and $\alpha_{H}\ge 0.2$) are ruled out by the 
present data (only one object in the Massardi galaxy sample can be considered
as a pure ADAF candidate), indicating that such radiatively 
inefficient accretion regimes must be very 
rare also at the sub-mJy levels probed by the ATESP 5~GHz sample.

\begin{table}[t]   
\centering         
\caption{Mean values and standard deviations of the low and high frequency 
spectral indices. }
\label{tab:means}           
\begin{tabular}{lrrrr}
\hline                                     
\hline                                     
Sample$^{(a)}$ & $\alpha_L$ & $\sigma$ & $\alpha_H$ & $\sigma$   \\ 
\hline
ATESP          &  $-0.18$ & (0.11)  & $-0.79$ & (0.16) \\
Massardi G  &  $-0.01$ & (0.08)  & $-0.35$ & (0.09) \\
ATESP + Massardi G far    &  $-0.08$  & (0.07)  & $-0.50$ & (0.08) \\
Massardi Q far     &  $0.21$  & (0.04)  & $-0.16$ & (0.03) \\ 
\hline 
\multicolumn{5}{l}{\footnotesize $^{(a)}$ 'Massardi G' and 'Massardi Q' 
respectively refer to galaxies and}\\
\multicolumn{5}{l}{\footnotesize \hspace{0.3cm} 
stellar identifications of the Massardi~\etal sample. 'Far' refers to }\\
\multicolumn{5}{l}{\footnotesize \hspace{0.3cm} $z>0.5$ sources 
(or unknown redshift for quasar candidates)}\\
\end{tabular}                                                
\end{table} 

For a more quantitative analysis of the two-colour diagram we 
statistically compared both the low-- and high--frequency spectral index 
distribution of the ATESP and the Massardi \etal galaxy samples by 
implementing the survival analysis methods described in Feigelson \& Nelson 
\cite{Fei85} and Isobe \etal \cite{Iso86}. This allows us to properly treat 
upper/lower limits in $\alpha_H/\alpha_L$ respectively.
Both $\alpha_L$ and $\alpha_H$ spectral index distributions are
consistent with being drawn 
from the same parent population. In fact this hypothesis is
rejected at only a $<2\sigma$ confidence level (Peto \& Prentice 
generalized Wilcoxon test, Isobe \& Feigelson \citealp{Iso90}). 
This is especially interesting when we consider 
that the ATESP and  the Massardi~\etal samples cover very different ranges in 
flux density (by two or three 
orders of magnitude).

A similar comparison can be made between galaxies and quasars. To improve our
statistics we consider the ATESP and the Massardi \etal galaxies as a single 
population, and we limit our analysis to $z>0.5$ sources, where most 
quasars (or quasar candidates) are located 
(see right panel of Fig.~\ref{fig:alfapanel}). 
In this case we find that the hypothesis that $z>0.5$ galaxies and quasars 
are extracted from the same parent population is rejected at a 
$>4.5\sigma$ level 
(Peto \& Prentice generalized Wilcoxon test, Isobe \& Feigelson 
\citealp{Iso90}). 
Table~\ref{tab:means} summarises the mean values for the samples discussed 
above (Kaplan-Meier estimator, Isobe \& Feigelson \citealp{Iso90}). 
We notice that the Massardi sub-sample of $z>0.5$ quasars and
quasar candidates (Massardi Q far) tend to be 
flatter or even more inverted than sources associated with galaxies, 
indicating that quasar-like objects are more commonly associated 
with compact sources and less with
the classical extended radio lobes in which the optically thin synchrotron 
radiation with steep radio spectra is dominant. 
On the other hand, no difference is seen in spectral steepening: 
if we compare the curvature $C=\alpha_H - \alpha_L$ (see e.g. Gregorini \etal 
\citealp{Gre84}), we find $C=-0.42 \pm 0.11$ 
(ATESP + Massardi G far), and $C=-0.37 \pm 0.05$ (Massardi Q far). 

Following Sadler \etal \cite{Sad06} we consider in more detail the 
distribution of sources in the four quadrants of the two-colour diagram, 
which represent different types of radio sources (defined earlier 
in this section). 
The distribution of objects over the four quadrants is shown in
Table~\ref{tab:twocolour} (where U stands for {\it Upturn}, I for {\it Inverted}, S for  {\it Steep} , P for 
{\it Peaked}).
Most ATESP sources are in the lower left quadrant (61\%, 
Table~\ref{tab:twocolour}), as for classical 
FRI and FRII radio galaxies. 
This may be due to the early type galaxy optical 
pre-selection, and in fact the entries of the Massardi et al. galaxy 
sample (M-G in Table~\ref{tab:twocolour}) are, considering the errors, 
practically the same as those of the ATESP sample, despite the very different 
flux range covered by the two samples.

In Table~\ref{tab:twocolour} we also compare the distribution of objects in 
three samples, all selected at high 
($\geq 20$ GHz) frequencies, where no optical pre-selection was applied
(Tucci~\etal~\citealp{Tuc08}; Sadler~\etal~\citealp{Sad06};
Massardi~\etal~\citealp{Mas08}).
There appear to be some trends with flux density, which we can attribute 
partly to selection effects
and partly to the different composition of the overall samples, 
in which quasar spectra may be more or less dominant, depending on
the limiting flux density. The {\it upturn} and 
{\it steep} spectra (i.e. the left hand side
of the two-colour diagram) tend to become more numerous at lower flux 
densities, while the inverse happens with {\it inverted} and {\it peaked} 
spectra (right hand side). These trends are individually significant only
at the $2\sigma$ level, again due to the small numbers involved, but appear 
to be quite consistent in all four quadrants. If we sum the $U$ and $S$ 
sources (those with $\alpha_L < 0$) and the $I$ and $P$ sources
($\alpha_L > 0$), the dependence on the flux density becomes quite obvious: 
whereas about 18 \% of the sources have a positive $\alpha_L$ ($I$ and $P$) 
at the flux density 
level of the Tucci \etal (\citealp{Tuc08}) sample, at the much higher flux
levels of the Massardi \etal sample this strongly increases to about 60 \%; 
this difference is definitely significant at about 4$\sigma$.
The fact that the number of rising spectra increases with flux limit was 
already noticed by Bolton \etal \cite{Bol04}.
No doubt these trends may be due to selection effects; however, the possible 
increase of {\it peaked} sources at higher
flux densities is harder to explain by selection effects only.

\begin{table}[t]   
\centering         
\caption{Spectral type statistics in five samples$^{(a)}$. }  
\label{tab:twocolour}           
\begin{tabular}{l|rrrr|rrrrrr}                       
\hline                                     
\hline                                     
    & \multicolumn{2}{c}{ATESP} & \multicolumn{2}{c|}{M-G} & \multicolumn{2}{c}{Tuc} & \multicolumn{2}{c}{Sad} & \multicolumn{2}{c}{M}    \\ 
    & \multicolumn{2}{c}{\scriptsize $>0.6$ mJy} & \multicolumn{2}{c|}{\scriptsize $>500$ mJy} & \multicolumn{2}{c}{\scriptsize $>20$ mJy} & \multicolumn{2}{c}{\scriptsize $>100$ mJy} & \multicolumn{2}{c}{\scriptsize $>500$ mJy}  \\ 
  & \multicolumn{4}{c|}{(percent)} & \multicolumn{6}{c}{(percent)} \\
\hline
U &   9 & $\pm$5  & 8 & $\pm$5 & 29 & $\pm$6 & 22 & $\pm$5 &  10 & $\pm$1 \\
I  & 13 &   7 & 12 &   6 &   7 &  3 & 20 &   4 & 28 &   5  \\
S & 61 & 14 & 47 & 11 & 53 &  7 & 34 &   6 & 30 &   5 \\
P & 17 &   8 & 33 & 11 & 11 &  3 & 24 &   5 & 32 &   5 \\  
\hline
\multicolumn{11}{l}{$^{(a)}$ ATESP 5~GHz sample (26 sources);  Massardi~\etal galaxy}\\
\multicolumn{11}{l}{\hspace{0.4truecm} sample (M-G, 36 sources);  Tucci~\etal\cite{Tuc08} sample (Tuc,}\\
\multicolumn{11}{l}{\hspace{0.4truecm} 102 sources);  Sadler~\etal\cite{Sad06} sample (Sad, 101 sources)}\\
\multicolumn{11}{l}{\hspace{0.4truecm} Massardi~\etal\cite{Mas08} overall sample (M, 218 sources)}\\
\end{tabular}                                                
\end{table} 

\section{The AGN component of the ATESP 5~GHz sample}
\label{sec:pdim}

\begin{figure*}[t]
\resizebox{18truecm}{!}{\includegraphics{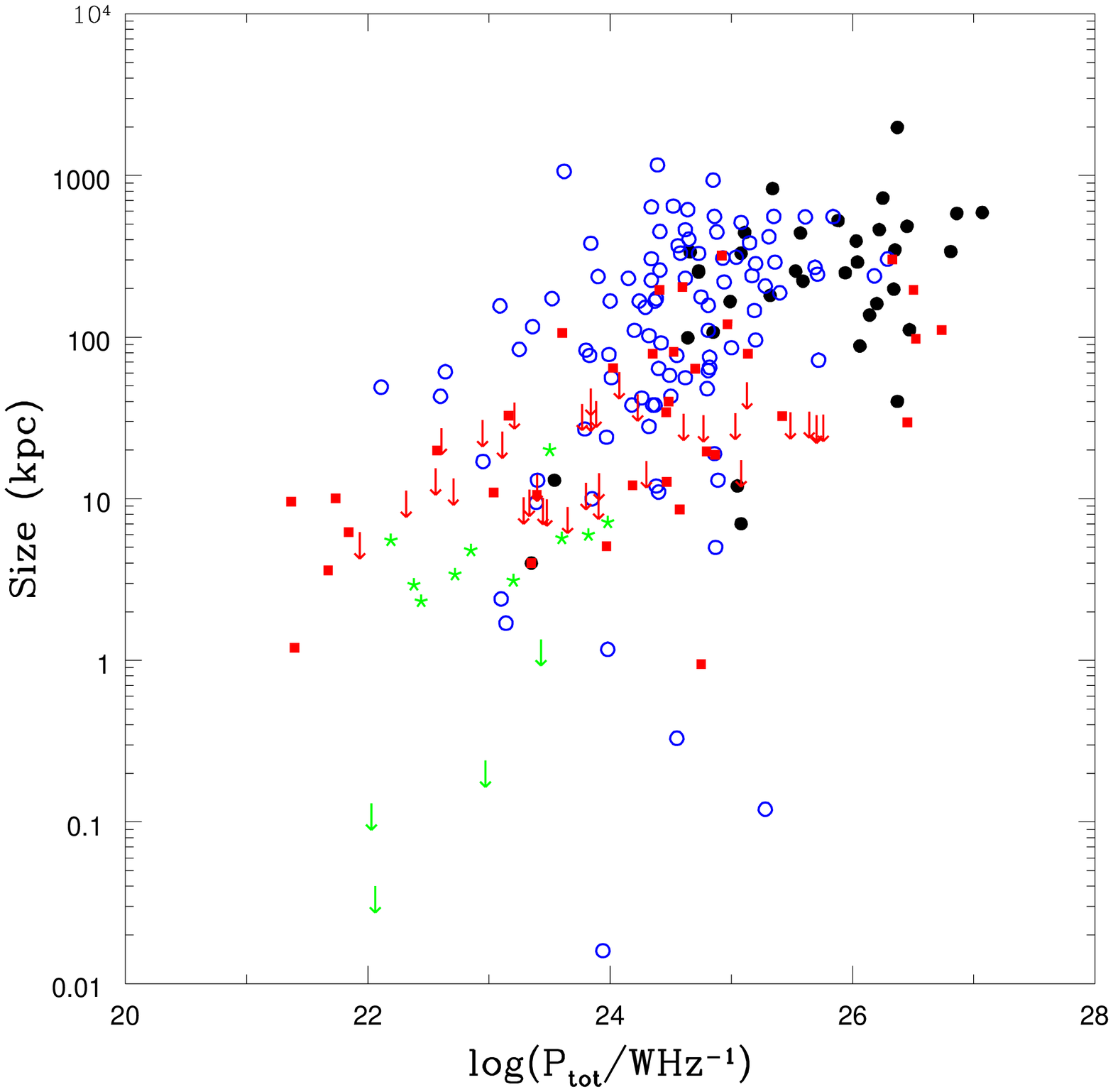}\includegraphics{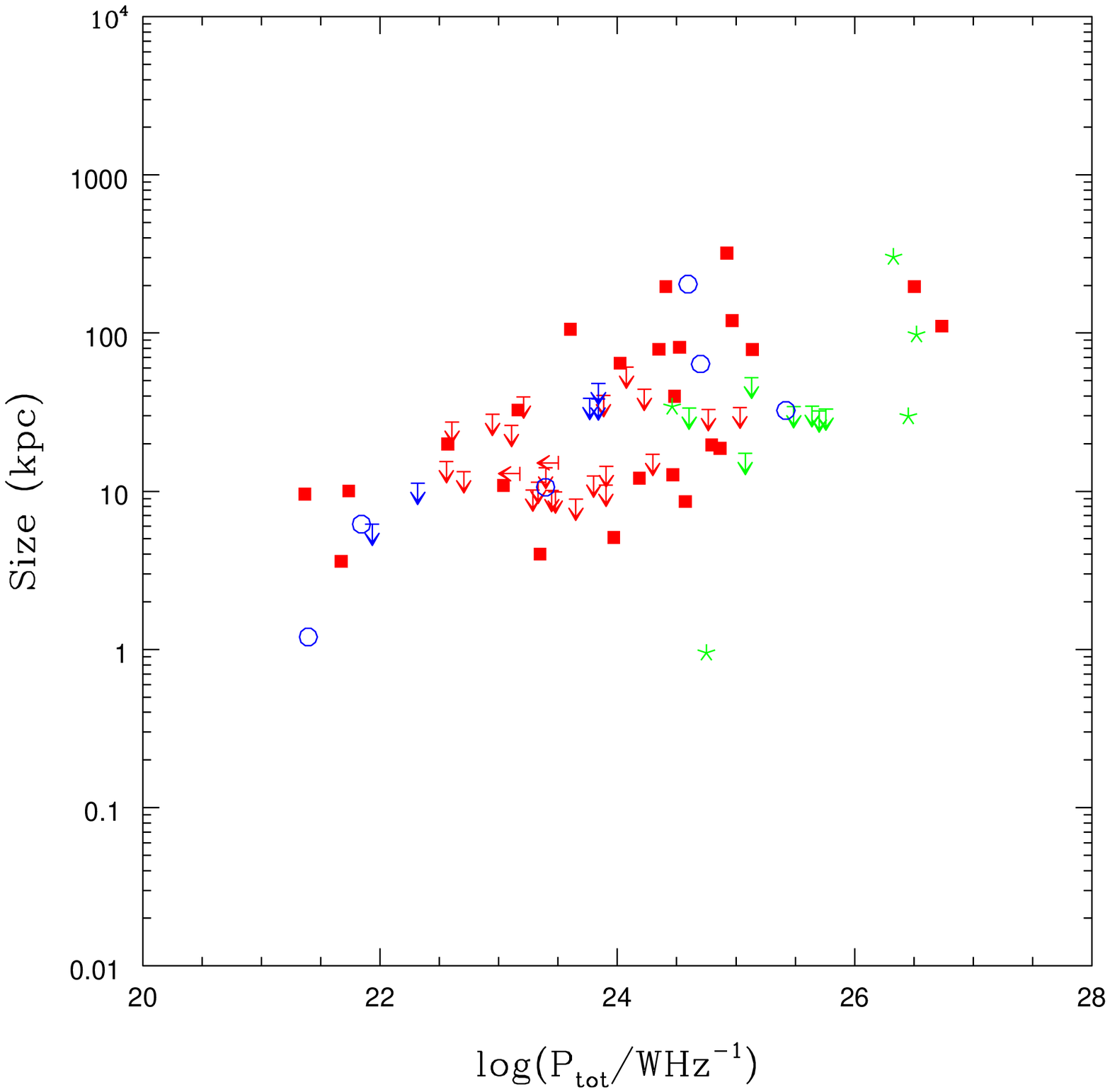}}
\caption[]{Linear size as a function of radio power (at 1.4 GHz). {\it 
Left panel:} 
comparison between four different samples: ATESP 5~GHz sample 
(red squares and upper limits), B2 catalogue (blue open circles), 
3C catalogue (black filled circles), radio-quiet QSO 
(green asterisks and upper limits). For Leipski \etal 
radio-quiet QSO sample 1.4 GHz luminosities were derived from tabulated 5 GHz 
values, assuming $\alpha=-0.7$. {\it Right panel:} comparison between different 
types of objects in the ATESP 5~GHz sample: sources associated to early type
galaxies (red squares and upper limits); sources associated to late type 
galaxies
(blue open circles and upper limits; among late type sources we may also
have some mis-classified narrow-line AGN); sources associated to 
broad-line AGNs, typically quasars (green asterisks and upper limits). }
\label{fig:pdim}
\end{figure*}

In Sect.~\ref{sec:results} we discussed the radio spectral 
properties of a representative sub-sample of ATESP radio sources 
associated to early type galaxies in comparison with some recent 
high-frequency radio surveys. 
The main results of such an analysis can be summarised as follows.
Most sources have spectra ranging from steep to moderately inverted,
which is consistent with them being jet-dominated systems. Strong spectral
similarities found between our and the much brighter 
Massardi~\etal galaxy sample may suggest that our radio sources are simply
lower luminosity counterparts of FRII radio galaxies.
No evidence for an
ADAF population was found, and {\it peaked} GPS-like sources seem to be 
preferentially found at higher flux densities than the ones probed by the 
ATESP. In addition quasar-like {\it upturn} sources are rare ($9\pm 5\%$, see
Table~\ref{tab:twocolour}). 
While this latter result is partly a consequence of the ATESP optical 
pre-selection, it argues against an obscured radio-quiet quasar population 
hidden among the ATESP 
early type galaxies. Nevertheless compact quasar-like objects may be present
among flat/moderately inverted jet-dominated sources.

To further explore this hypothesis and probe the AGN component of the ATESP 
sample at large we now compare linear sizes and radio powers
of all the ATESP sources with redshift determination (71 sources, 
see Table~\ref{tab:sample}), with the ones of the well-known B2 
(Colla \etal\citealp{Col75}; Fanti \etal\citealp{Fan78}) 
and 3CR (Laing \etal\citealp{Lai83}) radio source catalogues, and with sizes
and powers of  
a sample of optically-selected radio-quiet QSO 
(Leipski~\etal\citealp{Lei06}). B2 and 3C catalogues probe different ranges in 
radio luminosity, with the former dominated by FRI ($P<10^{25}$ W/Hz) 
and the latter by FRII ($P>10^{25}$ W/Hz)
radio galaxies. Such a comparison is shown in Fig.~\ref{fig:pdim} (left panel).
In the right-hand panel of Fig.~\ref{fig:pdim} the same plot is shown, where 
symbols now stand for different optical types of ATESP radio sources.

We first focus our analysis on ATESP resolved sources (red squares in left 
panel). Such sources have generally 
larger sizes than radio-quiet QSOs and seem to be more consistent with the B2 
radio galaxy size distribution. In particular they seem to confirm and 
extend to lower radio luminosity the
size-power relation found for B2 radio galaxies (de Ruiter 
\etal\citealp{Rui90}). When enlarging 
the analysis to unresolved sources (red upper limits in left panel), 
we notice a group of sources at $P>10^{25}$ W/Hz which clearly stand below
the B2/3C source size distribution. As shown in the right panel of 
Fig.~\ref{fig:pdim}, such sources are in fact 
associated to 
broad-line AGNs  (see green upper limits) and could be genuine compact 
sources, perhaps the radio-intermediate counterparts of radio-quiet QSOs. 
Unresolved sources with radio powers $P<10^{25}$ W/Hz are mostly associated
to early-type galaxies. In general the given upper limits do not allow us
to distinguish between B2 FRI-like and quasar-like sizes. Nevertheless 
in some cases we have $d\lsimeq 10$ kpc, making such sources more 
consistent with a radio-quiet scenario. Some of those sources 
may also be low power BL Lac, where the optical activity is very weak (see 
discussion in Sect.~\ref{sec:introduction}). From 
geometrical considerations we expect $<3-6\%$ (or $<2-4$) of such 
objects in the sample, assuming viewing angles $< 15$\dg$-20$\dg
(Zenzus \& Pearson \citealp{Zen87}).
Such fraction could be somewhat higher, due to possible sample 
contamination from boosted objects
belonging to a parent population fainter than the ATESP flux limit. 
For a more conclusive analysis of the size-power plot we need higher resolution
data for the ATESP sample. In particular VLBI imaging 
would allow us to better probe the low power BL Lac scenario. 
As a final remark it is 
worth noticing that ATESP source sizes may 
suffer from some underestimation. This because low surface brightness 
flux associated to radio jets could be missed by blind 
source Gaussian 
fitting procedures performed to construct the ATESP source catalogues (see 
Prandoni~\etal\citealp{Pra00b,Pra06}). 

\begin{figure*}[t]
\includegraphics[scale=0.9]{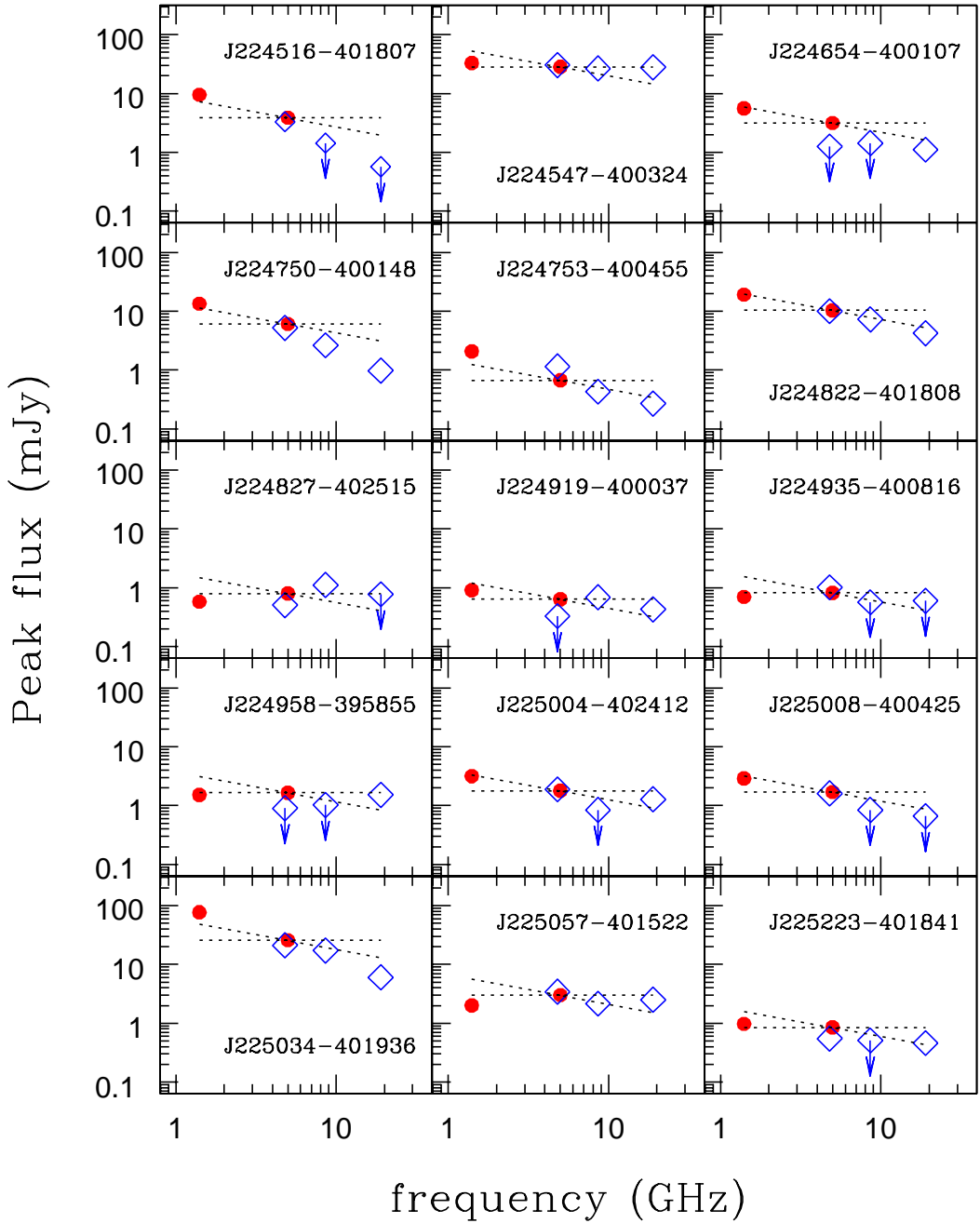}
\caption{Flux densities (in mJy) against frequency (in GHz) for the 
26 ATESP 5~GHz sources associated to early-type galaxies discussed in this work 
(see Table~\ref{tab:sample}).
Red filled circles represent the original 1.4 and 5 GHz points 
(Prandoni \etal \citealp{Pra00b} and Prandoni \etal \citealp{Pra06}), while 
blue diamonds represent the 2007 and 2008 observations discussed
in the present paper. No error bars are drawn, since these are typically 
smaller than symbols. Dotted lines indicate the expected slopes for 
$\alpha=0$ and $\alpha=-0.5$.}
\label{fig:fluxpanel}
\end{figure*}

\addtocounter{figure}{-1}
\begin{figure*}
\includegraphics[scale=0.9]{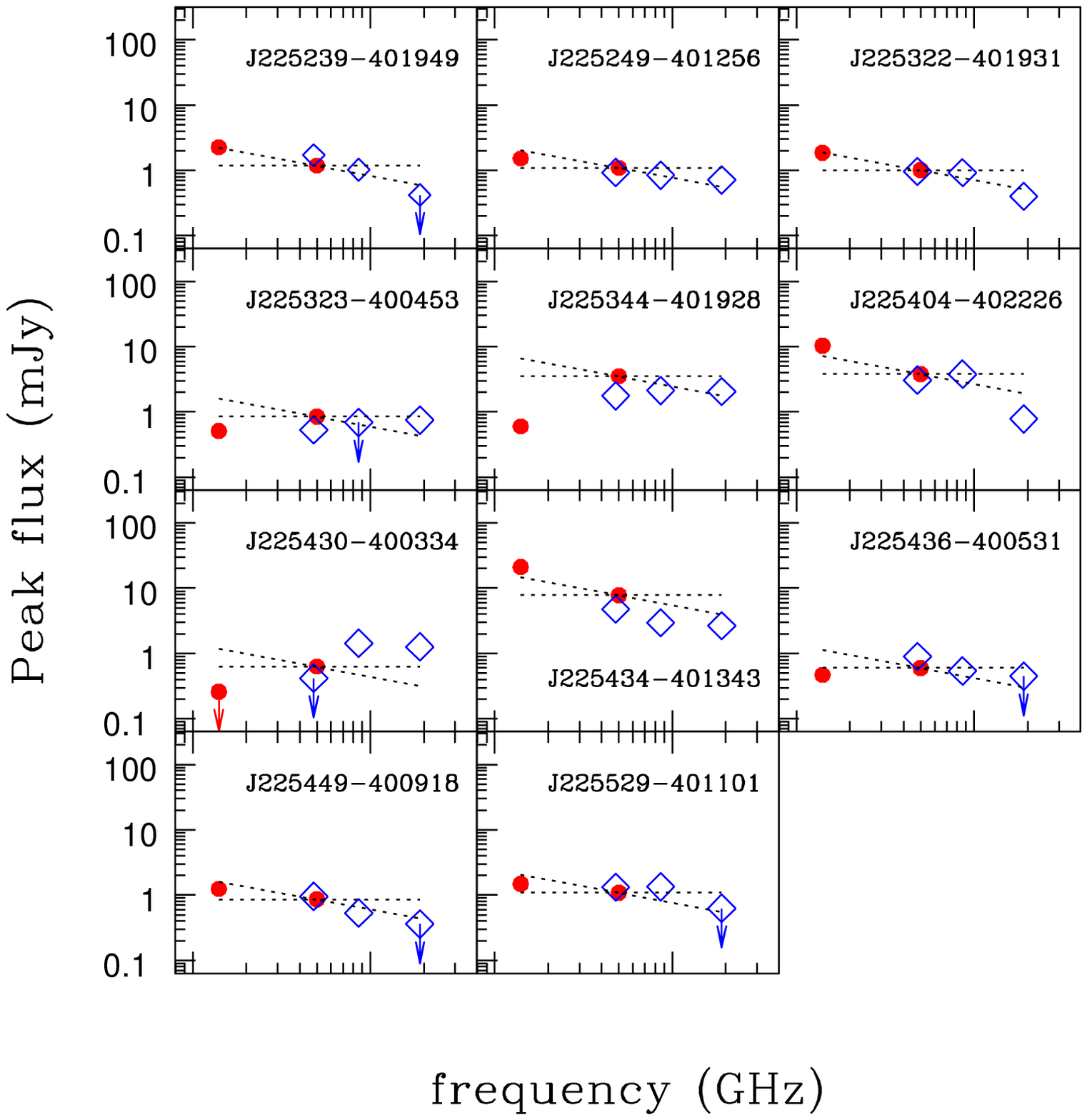}

\vspace{-3.0cm}
\caption{{\bf Continued}. Flux densities (in mJy) against frequency (in GHz) 
for the 26 ATESP 5~GHz sources associated to early-type galaxies discussed in 
this work (see Table~\ref{tab:sample}). }
\end{figure*}

\section{Summary}
\label{sec:sum}
 
We presented follow up quasi-simultaneous 
observations at 4.8, 8.6 and 19 GHz for a representative sub-sample of 
ATESP sources associated to early type galaxies (26 sources with $S>0.6$ mJy).
The new data were discussed together with the original 1.4 and 5 GHz
ATESP data (Prandoni \etal \citealp{Pra00b,Pra06}), 
with the aim of better understanding the radio spectral properties 
of such sources and ultimately the nature of AGNs at sub-mJy
flux levels. In particular we were interested in understanding whether the AGN
component of the sub--mJy population is more related to
efficiently accreting systems - like radio-intermediate/quiet quasars - or to
systems with low accretion rates - like e.g. FRI radio galaxies - or to
low radiative efficiency accretion flows - like e.g. ADAF.

From the analysis of the radio spectra, we found that our sources are 
most probably jet-dominated systems, with steeper sources dominated by
optically-thin synchrotron emission (typical of FRI and FRII radio galaxies)
and flatter or moderately-inverted sources dominated by 
optically-thick emission coming from the base of the jet. Pure ADAF models 
seem to be ruled out by the high frequency data, while ADAF+jet scenarios
could be consistent with flat/moderately inverted sources, 
but are not required to explain the data. 

By comparing our sample with some recent high frequency ($\gsimeq 20$ GHz) 
surveys we found strong spectral
similarities between our and the much brighter 
Massardi~\etal galaxy sample, possibly suggesting that 
our radio sources are mostly lower luminosity counterparts of the
Massardi~\etal (FRII) radio galaxies. In addition
{\it peaked} GPS-like sources seem to be 
preferentially found at higher flux densities than the ones probed by the 
ATESP sample, 
and quasar-like {\it upturn} sources are rare ($9\pm 5\%$). This latter 
result is partly due to the ATESP early type galaxy  
pre-selection, but argues against an obscured radio-quiet quasar population 
hidden among the ATESP 
early type galaxies. Nevertheless compact quasar-like objects may be present
among flat/moderately inverted jet-dominated sources.

To further investigate such a hypothesis we compared radio powers and 
linear sizes of the ATESP 5~GHz sample with the ones of two well-known
brighter radio source catalogues (B2 and 3CR) and with the sample of 
radio-quiet quasars studied by Leipski \etal\citep{Lei06}.
Unfortunately such a comparison suffers from the large number of unresolved 
ATESP sources (or the large number of size upper limits). 
A significant fraction of ATESP sources have sizes
consistent with the ones of B2 and 3CR radio galaxies. Such sources 
seem to confirm and extend to lower radio luminosity the
size-power relation found for B2 radio galaxies (de Ruiter 
\etal\citealp{Rui90}). On the other hand, a component of very compact 
$d<10$ kpc radio sources associated 
to early type galaxies seem to have sizes more consistent with the ones of
radio-quiet quasars. Among them we may have a few low power BL Lac. 
For a more conclusive analysis we need higher resolution (possibly VLBI)
data for the ATESP sample. 

\begin{acknowledgements}
The Australia Telescope Compact Array is part of the Australia Telescope 
which is funded by the Commonwealth of Australia for operation as a National 
Facility managed by CSIRO.
\end{acknowledgements}

\bibliographystyle{aa}

\begin{figure*}
\includegraphics[scale=0.8]{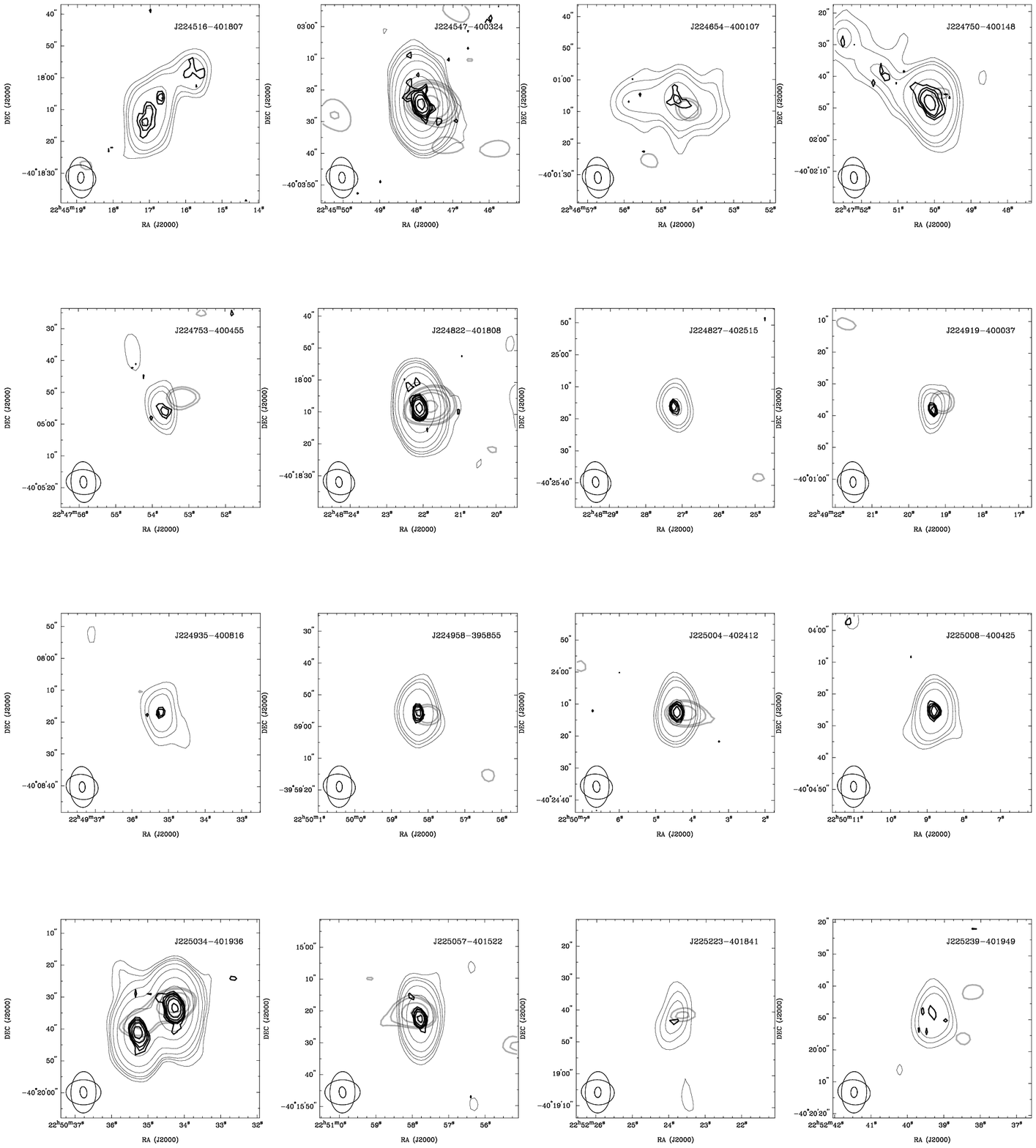}

\vspace{-2.0cm}
\caption{Contour images of the 26 ATESP 5~GHz sources identified with 
early-type galaxies, discussed in this work. Thin black contours 
represent low resolution 5~GHz data; superimposed are
high resolution 5~GHz contours (thick black lines) and low resolution K-band 
contours (thick gray lines).
Contour levels are 2.5 (only K-band), 3, 4.5, 6, 10, 20, 50 and 100 \% of 
the peak flux density.
}
\label{fig:overlays}
\end{figure*}

\addtocounter{figure}{-1}
\begin{figure*}
\includegraphics[scale=0.8]{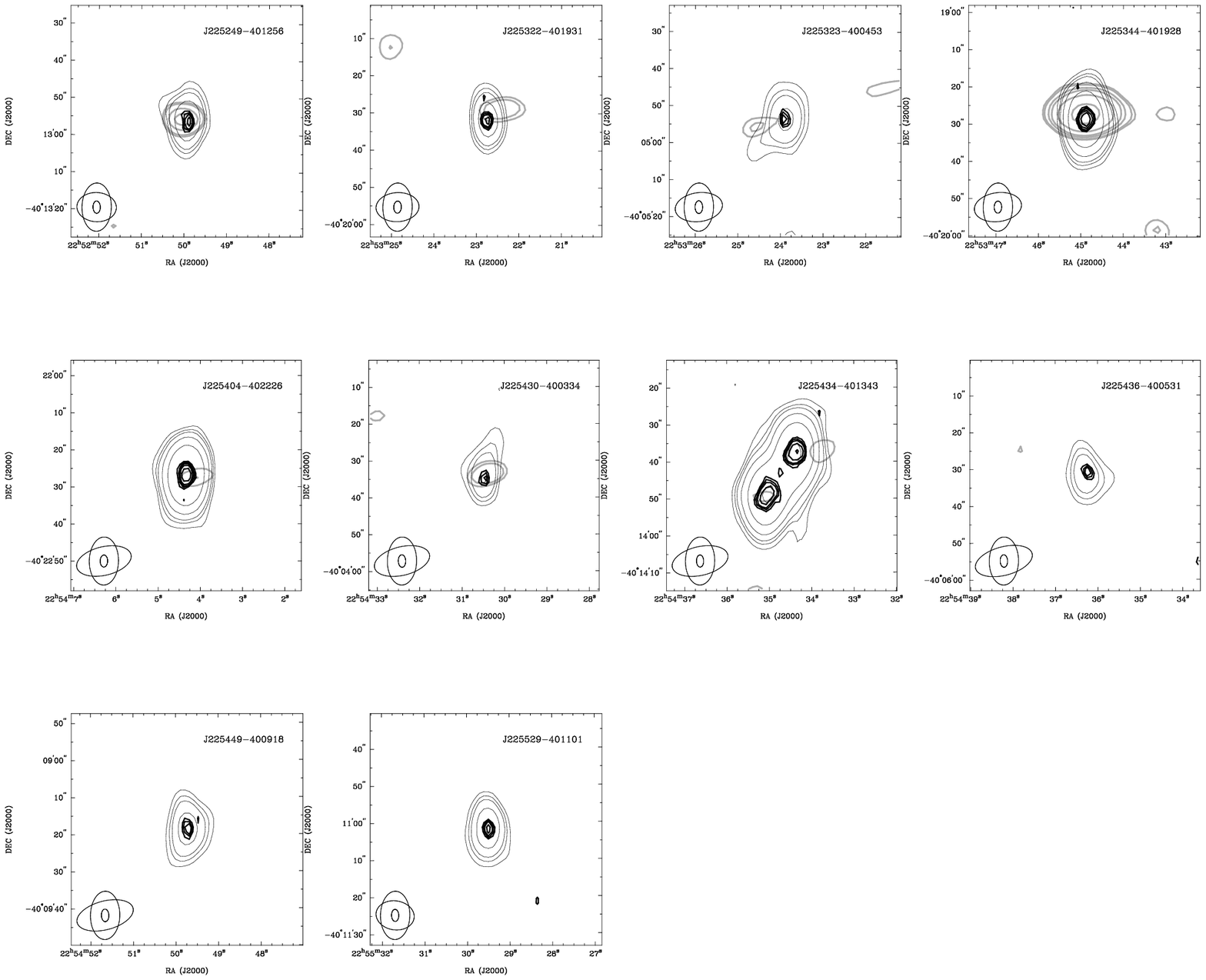}

\vspace{-7.0cm}
\caption{{\bf Continued.} Contour images of the 26 ATESP 5~GHz sources 
identified with early-type galaxies, discussed in this work.}
\end{figure*}

\end{document}